\documentclass[aps,preprint,showpacs,groupedaddress,floatfix]{revtex4}%
\usepackage{graphicx}
\usepackage{dcolumn}
\usepackage{bm}
\usepackage{amsmath}
\usepackage{amsfonts}
\usepackage{amssymb}%
\providecommand{\U}[1]{\protect\rule{.1in}{.1in}}
\providecommand{\U}[1]{\protect\rule{.1in}{.1in}}

\newcommand{\be}{\begin{equation}}
\newcommand{\ee}{\end{equation}}
\newcommand{\bea}{\begin{eqnarray}}
\newcommand{\eea}{\end{eqnarray}}

\newcommand{\ve}{\varepsilon}

\begin{document}
\title{Particle-vibration coupling within covariant density functional theory}
\author{E. Litvinova}
\affiliation{Physik-Department der Technischen Universit\"at M\"unchen, D-85748 Garching, Germany}
\affiliation{Institute of Physics and Power Engineering, 249033 Obninsk, Russia}
\author{P. Ring}
\affiliation{Physik-Department der Technischen Universit\"at M\"unchen, D-85748 Garching, Germany}
\author{V. Tselyaev}
\affiliation{Nuclear Physics Department, V. A. Fock Institute of Physics, St. Petersburg
State University, 198504, St. Petersburg, Russia}
\date{\today}

\begin{abstract}
Covariant density functional theory, which has so far been applied
only within the framework of static and time dependent mean field
theory is extended to include Particle-Vibration Coupling (PVC) in a
consistent way. Starting from a conventional energy functional we
calculate the low-lying collective vibrations in Relativistic Random
Phase Approximation (RRPA) and construct an energy dependent
self-energy for the Dyson equation. The resulting Bethe-Salpeter
equation in the particle-hole ($ph$) channel is solved in the Time
Blocking Approximation (TBA). No additional parameters are used and
double counting is avoided by a proper subtraction method. The same
energy functional, i.e. the same set of coupling constants, generates
the Dirac-Hartree single-particle spectrum, the static part of the
residual $ph$-interaction and the particle-phonon coupling vertices.
Therefore a fully consistent description of nuclear excited states
is developed. This method is applied for an investigation of damping
phenomena in the spherical nuclei with closed shells $^{208}$Pb and
$^{132}$Sn. Since the phonon coupling terms enrich the RRPA spectrum
with a multitude of $ph\otimes$phonon components a noticeable
fragmentation of the giant resonances is found, which is in full
agreement with experimental data and with results of the
semi-phenomenological non-relativistic approach.

\end{abstract}

\maketitle

\section{Introduction}

The theoretical analysis of nuclear dynamics by microscopic methods
remains undoubtedly a crucial issue of present-day nuclear physics.
Besides the merely theoretical interest it is stimulated by new
experimental opportunities for the study of the nuclear many-body
system. Along with new possibilities to investigate more precisely
nuclear systems being considered well known, the recent generation
of radioactive-beam facilities enables us to examine exotic systems
with extreme spin and isospin values. For a successful description
of nuclei close to the limits of stability a powerful and reliable
theory should be provided. It should be based on a consistent
treatment of both ground and excited states and it should allow for
predictions of nuclear properties in areas, which are hard or
impossible to access by future experiments.

Ab initio calculations and multi-configuration mixing within the shell-model
can, so far, only be applied in light nuclei. At present, for a universal
description of nuclei all over the periodic table, Density Functional Theory
(DFT) based on the mean-field concept provides a very reasonable concept. DFT
has been introduced in the sixties in atomic and molecular physics
\cite{HK.64,KS.65} and shortly after that in nuclear physics, where it has
been called density dependent Hartree-Fock theory \cite{VB.72,DG.80}. Today it
is widely used for all kinds of quantum mechanical many-body systems. Density
functional theory can, in principle, provide an exact description of the
many-body dynamics, if the exact density functional is known, but for such
systems as nuclei one is far from a microscopic derivation and one is forced
to determine the functional in a phenomenological way. Starting from basic
symmetries the parameters are adjusted to characteristic experimental data in
finite nuclei and nuclear matter.

One of the most successful schemes of this type is Covariant Density
Functional Theory (CDFT). It is based on Lorentz invariance connecting in a
consistent way the spin and spatial degrees of freedom in the nucleus.
Therefore, it needs only a relatively small number of parameters which are
adjusted to reproduce a set of bulk properties of spherical closed-shell
nuclei \cite{Rin.96,VALR.05}. A large variety of nuclear phenomena have been
described over the years within this kind of model: ground state properties of
finite spherical and deformed nuclei all over the periodic table \cite{GRT.90}
from light nuclei \cite{LVR.04a} to super-heavy elements \cite{LSRG.96,BBM.04}%
, from the neutron drip line, where halo phenomena are observed
\cite{MR.96}, to the proton drip line \cite{LVR.04} with nuclei
unstable against the emission of protons \cite{LVR.99}. Rotational
bands are treated within the relativistic cranking approximation
\cite{KR.89,ARK.00} and the Relativistic Random Phase Approximation
(RRPA) \cite{RMG.01} and quasi-particle RRPA \cite{PRN.03} have been
formulated as the small amplitude limit of time-dependent
Relativistic Mean-Field (RMF) models for a description of collective
and non-collective excitations. This method is successful, in
particular, for the description of the positions of giant resonances
and spin/isospin-excitations as the Isobaric Analog Resonance (IAR)
or the Gamow-Teller Resonance (GTR) \cite{PNV.04}. Recently it has
been also used for a theoretical interpretation of low-lying dipole
\cite{PRN.03} and quadrupole \cite{Ans.05} excitations.

With a few exceptions, where the quadrupole motion has been studied
within the relativistic Generator Coordinate Method (GCM)
\cite{NVR.06}, applications of covariant density functional theory
to the description of excited states are limited to relativistic
RPA, i.e. to configurations of $1p1h$-nature. None of these methods,
however, can be applied to the investigation of the coupling to more
complicated configurations, as it occurs for instance in the damping
phenomena causing the width of giant resonances. It is also known,
that in such theories, because of the low values of the effective
mass, the level density of the single-particle spectrum in the
vicinity of the Fermi surface is considerably reduced as compared to
the experimental data.

Based on the Fermi Liquid Theory (FLT) of Landau \cite{Lan.59} the Theory of
Finite Fermi Systems (TFFS) of Migdal \cite{Mig.67} is another very successful
method for the description of low-lying nuclear excitations \cite{RSp.74},
which is known since the fifties. It has several general properties in common
with density functional theory. First, both theories are know to be exact, at
least in principle, but in practice the parameters entering these theories
have to be determined in a phenomenological way by adjustment to experimental
data. Second, both theories are based in some sense on a single-particle
concept. Density functional theory uses the mean field concept with Slater
determinants in an effective single-particle potential as a vehicle to
introduce shell-effects in an exact theory and Fermi liquid theory is based on
the concept of quasi-particles obeying a Dyson equation, which are defined as
the basic excitations of the neighboring system with odd particle number.
Third, in practical applications both theories describe in the simplest
versions the nuclear excitations in the RPA approximation, i.e. by a linear
combination of $ph$-configurations in an average nuclear potential.

However, there are also essential differences between these two concepts.
First, in contrast to DFT, TFFS does not attempt to calculate the ground state
properties of the many-body system, but it describes the nuclear excitations
in terms of Landau quasi-particles and their interaction. Therefore the
experimental data used to fix the phenomenological parameters of the theories
are bulk properties of the ground state in the case of density functional
theory, and properties of single-particle excitations and of the collective
excitations in the case of finite Fermi systems theory. Second, in DFT the mean
field is determined in a self-consistent way and therefore the RPA spectrum
contains Goldstone modes at zero energy. This is usually not the case in TFFS
calculations, which are based on a non-relativistic shell-model potential,
whose parameters are adjusted to the experimental single-particle spectra.
Therefore there is no self-consistency in the RPA calculations of TFFS and the
Goldstone modes do not separate from the other modes. They are distributed
among the low-lying excitations. Third, modern versions of TFFS go much beyond
the mean field approximation. Using Green's function techniques the coupling
between quasi-particles and phonons has been investigated. Based on the
phonons calculated in the framework of the RPA one has included
particle-phonon coupling vertices and an energy dependence of the self-energy
in the Dyson equation \cite{RW.73,HS.76}. This leads also to an induced
interaction in the Bethe-Salpeter equation caused by the exchange of phonons
which also depends on the energy. The coupling of particles and phonons has
also been derived from Nuclear Field Theory (NFT) and its extensions
\cite{BM.75,BBB.77,BBB.85}. Many aspects of the coupling between the
quasi-particles and the collective vibrations have been investigated with
these techniques
\cite{BBBD.79,BB.81,CBGBB.92,CB.01,SBC.04,Tse.89,KTT.97,KST.04,Tse.05,LT.05}
as well as with other kinds of approaches beyond RPA \cite{Sol,DNSW.90} over the years.

The present work makes the attempt to find a combination of the basic ideas of
density functional theory and Landau-Migdal theory. The concept is similar to
earlier work in Refs. \cite{KS.79,KS.80,FTT.00}, where specific
non-relativistic energy functionals have been used to construct
Self-Consistent Theory of Finite Fermi Systems. We start here from a covariant
density functional $E[\rho]$ widely used in the literature. It is adjusted to
ground state properties of characteristic nuclei and, without any additional
parameter, it provides the necessary input of finite Fermi systems theory, such as the
mean field and the single-particle spectrum as well as an effective
interaction between the $ph$-configurations in terms of the second derivative
of the same energy $E[\rho]$ with respect to the density. Thus the
phenomenological input of Landau-Migdal theory is replaced by the results of
density functional theory. The same interaction is used to calculate the
vertices for particle-vibration coupling \cite{LR.06}. We then apply the
techniques of Landau-Migdal theory and its modern extensions to describe
vibrational coupling and complex configurations. In this way we obtain a fully
consistent description of the many-body dynamics.

Two essential approximations are used in this context: First, the
\textit{Time-Blocking Approximation} (TBA) blocks in a special
time-projection technique the $1p1h$-propagation through states
which have a more complex structure than $1p1h\otimes$phonon. The
nuclear response can then be explicitly calculated on the
$1p1h+1p1h\otimes$phonon level by summation of an infinite series of
Feynman's diagrams. This approximation has been introduced in
Ref.~\cite{Tse.89}. Initially it was called Method of Chronological
Decoupling of Diagrams (MCDD) and it is also discussed in recent
review articles \cite{KTT.97,KST.04}. Second, a special
\textit{subtraction technique} guarantees, that there is no double
counting between the additional correlations introduced by
particle-vibration coupling and the ground state correlations
already taken into account in the phenomenological density
functional. These two tools are essential for the success of the
present method. TBA introduces a consistent truncation scheme into
the Bethe-Salpeter equation and without it it would be hard to solve
the equations explicitly. The subtraction method is the essential
tool to connect density functional theory so far used only on the
level of mean field theory, i.e. on the RPA level, with the
extended Landau-Migdal theory, where complex configurations are
included through particle-vibration coupling.

As discussed above there are several versions of density functional
theory in nuclear system. In the present work we use covariant
density functional theory with the parameter set NL3 \cite{NL3}. A
first attempt in this direction has been carried out in
Ref.~\cite{LR.06}, where we treated the level density problem in the
framework of covariant particle-vibration coupling.

In Section II we discuss shortly the general formalism of covariant density
functional theory, we introduce the concept of the energy dependent
self-energy $\Sigma(\varepsilon)$ and the vertices of particle-vibration
coupling in the relativistic framework, and we discuss the response formalism
and the time blocking approximation for the response function. In Section III
we show, how to solve the resulting integral equations in detail and present
recent numerical applications for the spreading width of the doubly magic
nuclei $^{208}$Pb and $^{132}$Sn and in Section IV we include a brief summary
and an outlook for future applications. Appendices A-D provide some more
details for the formalism used in the calculations.

\section{Formalism}

\subsection{The relativistic mean-field approach}

In relativistic mean-field theory the nucleus is considered as a
system of nucleons interacting by the exchange of effective mesons,
which characterize the quantum numbers of the various fields.
Conventional RMF theory uses $\sigma$, $\omega$ and $\rho$-mesons
and the photon as a minimal set of bosons which is necessary to
generate realistic nuclear fields \cite{Wal.74,SW.86,Rin.96}. This
concept is expressed by the following Lagrangian density depending
on the nucleon spinor $\psi$ as well as on the mesons $\sigma$,
$\omega^{\mu}$, ${\vec{\rho}}^{\mu}$ and the electromagnetic fields
$A^{\mu}$ :
\begin{align}
\mathcal{L}  &  ={\bar{\psi}}(i\gamma^{\mu}\partial_{\mu}-m)\psi+\frac{1}%
{2}\partial^{\mu}\sigma\partial_{\mu}\sigma-\frac{1}{2}m_{\sigma}^{2}%
\sigma^{2}\nonumber\\
&  -\frac{1}{4}\Omega^{\mu\nu}\Omega_{\mu\nu}+\frac{1}{2}m_{\omega}^{2}%
\omega^{\mu}\omega_{\mu}-\frac{1}{4}{\vec{R}}^{\mu\nu}{\vec{R}}_{\mu\nu}%
+\frac{1}{2}m_{\rho}^{2}{\vec{\rho}}^{\ \mu}{\vec{\rho}}_{\mu}-\frac{1}%
{4}F^{\mu\nu}F_{\mu\nu}\nonumber\\
&  -{\bar{\psi}}\Bigl({\Gamma}_{\sigma}\sigma+\Gamma_{\omega}^{\mu}\omega
_{\mu}+{\vec{\Gamma}}_{\rho}^{\ \mu}{\vec{\rho}}_{\mu}+\Gamma_{e}^{\mu}A_{\mu
}\Bigr)\psi. \label{Lagr}%
\end{align}
Here $m$ is the bare nucleon mass, $m_{\sigma}$, $m_{\omega}$ and $m_{\rho}$
are the corresponding meson masses, and $\Omega_{\mu\nu}$, ${\vec{R}}_{\mu\nu
},F_{\mu\nu}$ are field tensors:
\begin{align}
\Omega_{\mu\nu}  &  =\partial_{\mu}\omega_{\nu}-\partial_{\nu}\omega_{\mu
}\ ,\nonumber\\
{\vec{R}}_{\mu\nu}  &  =\partial_{\mu}{\vec{\rho}}_{\nu}-\partial_{\nu}%
{\vec{\rho}}_{\mu}\ ,\nonumber\\
F_{\mu\nu}  &  =\partial_{\mu}A_{\nu}-\partial_{\nu}A_{\mu}.
\end{align}
As usual arrows denote isovectors and boldface symbols will be used
for vectors in ordinary space. The vertices $\Gamma$ entering the
interaction term of Eq.~(\ref{Lagr}) read
\begin{equation}
\Gamma_{\sigma}=g_{\sigma},\ \ \ \ \Gamma_{\omega}^{\mu}=g_{\omega}\gamma
^{\mu},\ \ \ \ {\vec{\Gamma}}_{\rho}^{\ \mu}=g_{\rho}{\vec{\tau}}\gamma^{\mu
},\ \ \ \ \Gamma_{e}^{\mu}=e\frac{(1-\tau_{3})}{2}\gamma^{\mu}, \label{gammas}%
\end{equation}
where $g_{\sigma}$, $g_{\omega}$, $g_{\rho}$ and $e$ are the
corresponding coupling constants. For the sake of simplicity, we use
in the following discussions the linear version (\ref{Lagr}) of the
Lagrangian. However, it is well known that one needs an additional
density dependence in order to obtain a reliable description of
nuclear surface properties. It is either expressed by a non-linear
self-interaction between the mesons or by an explicit density
dependence of the coupling constants $g_{\sigma}$, $g_{\omega}$, and
$g_{\rho }$. In all the numerical applications we use such
modifications. They are discussed in detail in Appendix~\ref{nls}.

The classical variation principle being applied to the Lagrangian density
(\ref{Lagr}) leads to the following system of coupled equations of motion for
the Fermions
\begin{equation}
(i\gamma^{\mu}\partial_{\mu}-m-\sum\limits_{m}\Gamma_{m}\phi_{m}%
(\mathbf{r},t))\psi_{i}(\mathbf{r},t)=0 \label{rmf1}%
\end{equation}
and the mesons:
\begin{equation}
(\Box+m_{m}^{2})\phi_{m}(\mathbf{r},t))=\mp\sum\limits_{i}{\bar{\psi}}%
_{i}(\mathbf{r},t)\Gamma_{m}{\psi}_{i}(\mathbf{r},t). \label{rmf2}%
\end{equation}
Here the index $i$ numerates the nucleons and the index $m=\{\sigma
,\omega,\rho,e\}$ runs over the set of meson fields
$\phi_{m}=\{\sigma ,\omega^{\mu},{\vec{\rho}}^{\mu},A^{\mu}\}$.
$m_{m}$ are the corresponding masses and $\Gamma_{m}$ are the
vertices (\ref{gammas}). The $-$sign in Eq.~(\ref{rmf2}) holds for
scalar fields and the $+$sign for vector fields. The set of
Eqs.~(\ref{rmf1}) and (\ref{rmf2}) defines the standard RMF model.
It implies the following four approximations:

i) only the motion of the nucleons is quantized, the meson and electromagnetic
degrees of freedom are described by \textit{classical fields}. The nucleons
move independently in these classical fields, i.e. residual two-body
correlations are disregarded and the many-nucleon wave function is a Slater
determinant at all times.

ii) in time-dependent applications we use the \textit{instantaneous
approximation}, i.e. the time derivatives $\partial_{t}^{2}$ in the
Klein-Gordon equations and consequently the retardation effects for the meson
fields are neglected although the Dirac spinors $\psi_{i}$ as well as the
fields $\phi_{m}$ are functions of coordinates and time. This is justified by
the relatively large meson masses and the corresponding short range of the
meson exchange interaction.

iii) \textit{Fock terms} are neglected. Because of the short range character
of the meson exchange forces they behave similar to zero range forces, where
the Fock terms have the same form as the direct terms and contribute only in a
change of the relevant strength parameters. Since our strength parameters are
adjusted to experiment, Fock terms for the heavy mesons are implicitly taken
into account to a large extend.

iv) the \textit{no-sea approximation} means, that nucleon states in
the Dirac sea with negative energies do not contribute to the
densities and currents: the sum over $i$ in the Eq.~(\ref{rmf2})
includes only occupied levels with positive energy in the Fermi sea.
This means we do not consider vacuum polarization explicitly. Since
our parameters are adjusted to experimental data such effects are
taken into account in an implicit way.

\subsection{The energy-dependent nucleon self-energy}

Let us now consider in detail the Dyson equation, which describes the motion
of a single nucleon in the nuclear interior:
\begin{equation}
(i\gamma^{\mu}\partial_{\mu}-m-\Sigma)\psi_{i}=0.
\end{equation}
Here the total self-energy $\Sigma$ contains all the forces acting on the
nucleon. As long as one stays within RMF approach the nucleon self-energy
contains only static and local contributions from the mesons and from the
electromagnetic fields.

When we go beyond the mean-field approximation, we have to take into account
that in the general case the full self-energy $\Sigma$ is non-local in the
space coordinates as well as in time. This non-locality means that its Fourier
transform has both momentum and energy dependence. Let us now decompose the
total self-energy matrix into two components, a static local and an energy
dependent non-local term:
\begin{equation}
\Sigma(\mathbf{r},\mathbf{r^{\prime}};\varepsilon)={\tilde{\Sigma}}%
(\mathbf{r})\delta(\mathbf{r}-\mathbf{r^{\prime}})+\Sigma^{e}(\mathbf{r}%
,\mathbf{r^{\prime}};\varepsilon), \label{sig}%
\end{equation}
where
\begin{equation}
{\tilde{\Sigma}}(\mathbf{r})=\sum\limits_{m}\Gamma_{m}\phi_{m}(\mathbf{r})
\end{equation}
is the static part known from RMF theory in Eq. (\ref{rmf1}). The
upper index "e" in $\Sigma^{e}$ indicates the energy dependence. The
Dyson equation for single-particle Green's function reads:
\begin{equation}
(\varepsilon-h^{\mathcal{D}}-\beta\Sigma^{e}(\varepsilon))G(\varepsilon
)=\beta, \label{fg}%
\end{equation}
where $h^{\mathcal{D}}$ denotes the Dirac hamiltonian in Eq.
(\ref{rmf1}) with the energy-independent mean field:
\begin{equation}
h^{\mathcal{D}}={\mbox{\boldmath $\alpha$}}\mathbf{p}+\beta(m+{\tilde{\Sigma}%
}_{s})+{\tilde{\Sigma}}_{0},\ \ \ \ \ {\mbox{\boldmath $\alpha$}}%
=\beta{\mbox{\boldmath $\gamma$}},\ \ \ \beta=\gamma^{0}. \label{hd}%
\end{equation}
Most of the static calculations are done for cases, where time-reversal
symmetry is valid. In all these cases there are no currents in the nucleus
and, thus, only scalar ${\tilde{\Sigma}}_{s}$ and time-like ${\tilde{\Sigma}%
}_{0}$ components of ${\tilde{\Sigma}}$ survive in the static Dirac
hamiltonian of Eq.~(\ref{hd}). In the shell-model Dirac basis
$\{|\psi _{k}\rangle\}$ which diagonalizes the energy-independent
part of the Dirac equation
\begin{equation}
h^{\mathcal{D}}|\psi_{k}\rangle=\varepsilon_{k}|\psi_{k}\rangle
\label{dirac-basis}%
\end{equation}
one can rewrite Eq.~(\ref{fg}) as follows:
\begin{equation}
\sum\limits_{l}\bigl\{(\varepsilon-\varepsilon_{k})\delta_{kl}-\Sigma_{kl}%
^{e}(\varepsilon)\bigr\}G_{lk^{\prime}}(\varepsilon)=\delta_{kk^{\prime}},
\label{fg1}%
\end{equation}
where
\begin{equation}
\Sigma_{kl}^{e}(\varepsilon)=\int d^{3}rd^{3}{r}^{\prime}~\bar{\psi}%
_{k}(\mbox{\boldmath $r$})\Sigma^{e}({\mbox{\boldmath $r$}}%
,{\mbox{\boldmath $r$}^{\prime}};\varepsilon)\psi_{l}%
({\mbox{\boldmath $r$}^{\prime}})\,, \label{sigmae}%
\end{equation}%
\begin{equation}
G_{kl}(\varepsilon)=\int d^{3}rd^{3}{r}^{\prime}~\bar{\psi}_{k}%
(\mbox{\boldmath $r$})\,\beta
G({\mbox{\boldmath $r$}},{\mbox{\boldmath $r$}^{\prime}};\varepsilon)
\beta\,\psi_{l}({\mbox{\boldmath $r$}^{\prime}})\,. \label{gfme}%
\end{equation}
The letter indices $k,k^{\prime},l$ denote full sets of the single-particle
quantum numbers. In the present work we consider spherical symmetry, so that
the spinor $|\psi_{k}\rangle$ is characterized by the set $k=\{(k),m_{k}\}$
and $(k)=\{n_{k},j_{k},\pi_{k},{\tau}_{k}\}$ with the radial quantum number
$n_{k}$, angular momentum quantum numbers $j_{k},m_{k}$, parity $\pi_{k}$ and
isospin ${\tau}_{k}$:
\begin{equation}
\psi_{k}(\mathbf{r},s,t)=\left(
\begin{array}
[c]{c}%
f_{(k)}(r)\Phi_{l_{k}j_{k}m_{k}}(\vartheta,\varphi,s)\\
ig_{(k)}(r)\Phi_{{\tilde{l}}_{k}j_{k}m_{k}}(\vartheta,\varphi,s)
\end{array}
\right)  \chi_{{\tau}_{k}}(t),
\end{equation}
where $s$ and $t$ are the coordinates for spin and isospin. The orbital
angular momenta $l_{k}$ and $\tilde{l}_{k}$ of the large and small components
are determined by the parity of the state $k$:
\begin{equation}
\left\{
\begin{array}
[c]{ccc}%
l_{k}=j_{k}+\frac{1}{2}, & {\tilde{l}}_{k}=j_{k}-\frac{1}{2} &
\mbox{for}\ \ \pi_{k}=(-1)^{j_{k}+\frac{1}{2}}\\
l_{k}=j_{k}-\frac{1}{2}, & {\tilde{l}}_{k}=j_{k}+\frac{1}{2} &
\mbox{for}\ \ \pi_{k}=(-1)^{j_{k}-\frac{1}{2}},
\end{array}
\right.
\end{equation}
$f_{(k)}(r)$ and $g_{(k)}(r)$ are radial wave functions and $\Phi
_{ljm}(\vartheta,\varphi,s)$ is a two-dimensional spinor:
\begin{equation}
\Phi_{ljm}(\vartheta,\varphi,s)=\sum\limits_{m_{s}m_{l}}({\frac
{{\scriptstyle1}}{{\scriptstyle2}}}m_{s}lm_{l}|jm)Y_{lm_{l}}(\vartheta
,\varphi)\chi_{m_{s}}(s).
\end{equation}

\begin{figure}[ptb]
\begin{center}
\includegraphics*[scale=0.75]{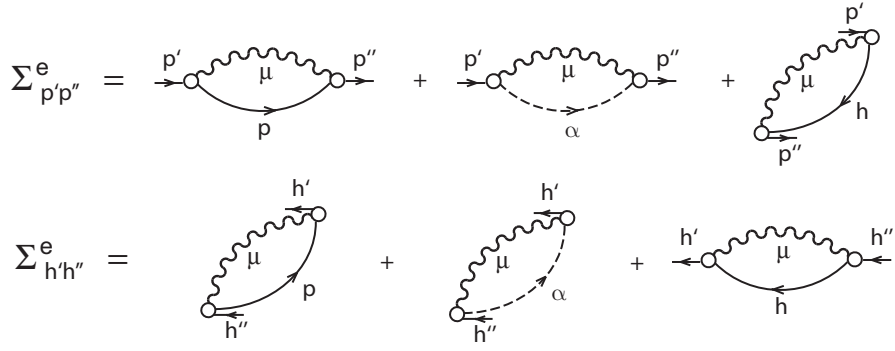}
\end{center}
\caption{The particle $\Sigma^{e}_{p^{\prime}p^{\prime\prime}}$ and the hole
$\Sigma^{e}_{h^{\prime}h^{\prime\prime}}$ components of the relativistic mass
operator in the graphical representation. Solid and dashed lines with arrows
denote one-body propagators for particle ($p$), hole ($h$), and antiparticle
($\alpha$) states. Wavy lines denote phonon ($\mu$) propagators, empty circles
are the particle-phonon coupling amplitudes $\gamma^{\mu}$. See text for the
detailed explanation.}%
\label{f0}%
\end{figure}

As in Ref.~\cite{LR.06} we use the particle-phonon coupling model
for the energy dependent part of the self-energy
$\Sigma^{e}(\varepsilon)$. In the Dirac basis of the spinors
$|\psi_{k}\rangle$ defined in Eq.~(\ref{dirac-basis}) it has the
form
\begin{equation}
\Sigma_{kl}^{e}(\varepsilon)=\sum\limits_{\mu,n}\frac{\gamma_{kn}^{\mu
(\sigma_{n})}\,\gamma_{ln}^{\mu(\sigma_{n})\ast}}{\varepsilon-\varepsilon
_{n}-\sigma_{n}(\Omega^{\mu}-i\eta)},\qquad\gamma_{kn}^{\mu(\sigma)}%
=\delta_{\sigma,+1}^{\vphantom{A}}\gamma_{kn}^{\mu}+\delta_{\sigma
,-1}^{\vphantom{A}}\gamma_{nk}^{\mu\ast}\ , \label{mo}%
\end{equation}
where $\eta\rightarrow+0$. In Fig.~\ref{f0} we show a graphical
representation for $\Sigma_{kl}^{e}(\varepsilon)$. The index $n$
runs over all single-particle states in this basis. We distinguish
occupied states in the Fermi sea $|h\rangle$ (hole states) with
$\sigma_{h}=-1$, unoccupied states above the Fermi level $|p\rangle$
(particle states) with $\sigma_{p}=+1$, and states in the Dirac sea
$|\alpha\rangle$ (anti-particle states) with negative energies. Due
to the no-sea approximation the orbits with negative energies
$|\alpha\rangle$ are left unoccupied and therefore we have
$\sigma_{\alpha }=+1$. However, as it has been shown in
Ref.~\cite{LR.06} the numerical contribution of the diagrams with
intermediate states $\alpha$ with negative energy is very small due
to the large energy denominators in the corresponding terms of the
self-energy (\ref{mo}). The index $\mu$ in Eq.~(\ref{mo}) labels the
various phonons taken into account. $\Omega^{\mu}$ is their
frequency, ${\rho}^{\mu}$ is their transition density and the
phonon vertices $\gamma_{kl}^{\mu}$ determine their coupling to the
nucleons:
\begin{equation}
\gamma_{kl}^{\mu}=\sum\limits_{k^{\prime}l^{\prime}}\tilde{V}_{kl^{\prime
},lk^{\prime}}^{\vphantom{A}}{\rho}_{k^{\prime}l^{\prime}}^{\mu}.
\label{phonon}%
\end{equation}
$\tilde{V}_{kl^{\prime},lk^{\prime}}$ denotes the relativistic matrix element
of the residual interaction. It is a obtained as functional derivative of the
relativistic mean field with respect to relativistic density matrix ${\rho}$
\begin{equation}
\tilde{V}_{kl^{\prime},lk^{\prime}}=\frac{\delta{\tilde{\Sigma}}_{l^{\prime
}k^{\prime}}}{\delta\rho_{lk}}. \label{static-interaction}%
\end{equation}
We use the linearized version of the model which assumes that $\rho^{\mu}$
is not influenced by the particle-phonon coupling and can be
computed within the relativistic RPA approximation. As long as we
neglect the non-linear self-couplings of the mesons or the density
dependence of the coupling constants $g_{m}$ in Eq.~(\ref{gammas})
the matrix elements
$\tilde{V}_{kl^{\prime},lk^{\prime}}^{\vphantom{A}}$ in
Eq.~(\ref{phonon}) have simple Yukawa form. However, such effects
have to be taken into account to guarantee full self-consistency of
the RRPA equations and we find more complicated expressions, which
are discussed in detail in Appendix~\ref{nls}. In the applications
investigated in section III we use the non-linear parameter set NL3
\cite{NL3} for the solution of the static Dirac equation
(\ref{dirac-basis}) as well as for the calculation of the matrix
elements (\ref{static-interaction}).

\subsection{The response function}

\begin{figure}[ptb]
\begin{center}
\includegraphics*[scale=0.75]{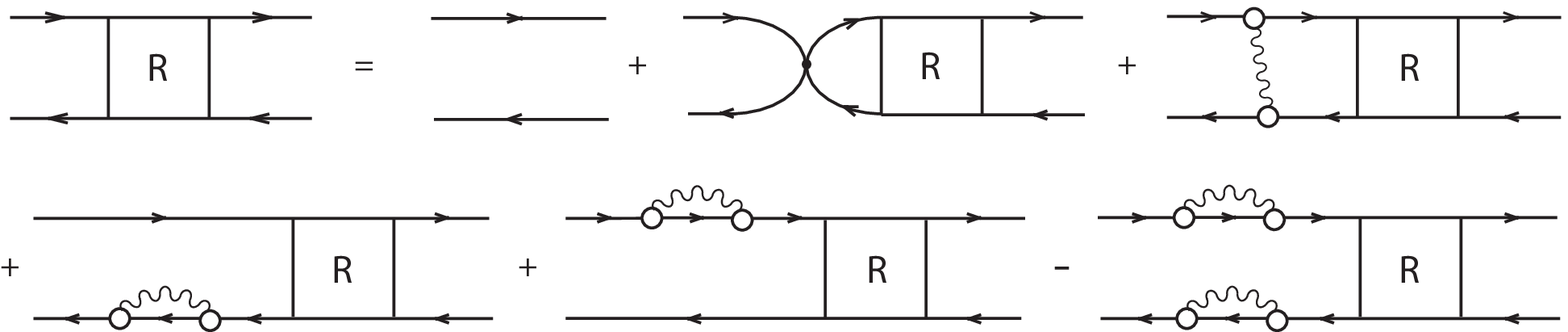}
\end{center}
\caption{Bethe-Salpeter equation for the $ph$-response function $R$
in graphical representation. Details are given in Fig.~\ref{f0} and
the small black circle means the static part of the residual
$ph$-interaction
(\ref{static-interaction}). }%
\label{f1}%
\end{figure}

Nuclear dynamics of an even-even nucleus in a weak external
%electromagnetic
field is described by the linear response function $R(14,23)$ which is the
solution of the Bethe-Salpeter equation (BSE) in the $ph$ channel. In the
beginning it is convenient to consider this equation in the time
representation. Let us include the time variable into the single-particle
indices setting $1=\{k_{1},t_{1}\}$. In this notation the Bethe-Salpeter
equation BSE for the response function $R$ reads:
\begin{equation}
R(14,23)=G(1,3)G(4,2)-i\sum\limits_{5678}G(1,5)G(6,2)V(58,67)R(74,83),
\label{bse}%
\end{equation}
where the summation over the number indices $1$, $2,\dots$ implies integration
over the respective time variables. The function $G$ is the exact
single-particle Green's function, and $V$ is the amplitude of the effective
interaction irreducible in the $ph$-channel. This amplitude is determined as a
variational derivative of the full self-energy $\Sigma$ with respect to the
exact single-particle Green's function:
\begin{equation}
V(14,23)=i\frac{\delta\Sigma(4,3)}{\delta G(2,1)}. \label{uampl}%
\end{equation}
Introducing the free response $R^{0}(14,23)=G(1,3)G(4,2)$ the Bethe-Salpeter
equation (\ref{bse}) can be written in a shorthand notation as
\begin{equation}
R=R^{0}-iR^{0}VR.
\end{equation}
For the sake of simplicity, we will use this shorthand notation
frequently in the following discussions. Since the self-energy in
Eq.~(\ref{sig}) has two parts $\Sigma=\tilde{\Sigma}+\Sigma^{e}$,
the effective interaction $V$ in Eq.~(\ref{bse}) is a sum of the
static RMF interaction $\tilde{V}$ and time-dependent terms $V^{e}$:
\begin{equation}
V=\tilde{V}+V^{e},
\end{equation}
where (with $t_{12}=t_{1}-t_{2}$)
\begin{equation}
\tilde{V}(14,23)=\tilde{V}_{k_{1}k_{4},k_{2}k_{3}}\delta(t_{31})\delta
(t_{21})\delta(t_{34}) \label{V-static}%
\end{equation}
is the static part of the interaction (see Eq.~(\ref{static-interaction})) and%
\begin{equation}
V^{e}(14,23)=i\frac{\delta{\Sigma^{e}(4,3)}}{\delta G(2,1)} \label{dcons}%
\end{equation}
contains the energy dependence. In the space of the Dirac basis
\ (\ref{dirac-basis}) the amplitude $V^{e}$ has the form:
\begin{equation}
V_{kl^{\prime},lk^{\prime}}^{e}(\omega,\varepsilon,\varepsilon^{\prime}%
)=\sum\limits_{\mu,\sigma}\frac{\sigma\gamma_{k^{\prime}k}^{\mu(\sigma)\ast
}\gamma_{l^{\prime}l}^{\mu(\sigma)}}{\varepsilon-\varepsilon^{\prime}%
+\sigma(\Omega^{\mu}-i\eta)}. \label{ueampl}%
\end{equation}
In order to make the Bethe-Salpeter equation (\ref{bse})
more convenient for the further analysis
we eliminate the exact Green's function $G$ and rewrite it
in terms of the mean field Green's function $\tilde{G}$:
\begin{equation}
{\tilde{G}}(1,2)=-i\sigma_{k_{1}}\delta_{k_{1}k_{2}}\theta(\sigma_{k_{1}}%
\tau)e^{-i\varepsilon_{k_{1}}\tau},\ \ \ \ \ \tau=t_{1}-t_{2},
\end{equation}
where $\theta(t)$ is the Heaviside function. After a Fourier transformation in
time this reads%
\begin{equation}
{\tilde{G}}_{k_{1}k_{2}}(\varepsilon)=\frac{\delta_{k_{1}k_{2}}}%
{\varepsilon-\varepsilon_{k_{1}}+i\sigma_{k_{1}}\eta}. \label{G-tilde}%
\end{equation}
Using the connection between $\tilde{G}$ and $G$ in the Nambu form
\begin{equation}
{\tilde{G}}^{-1}(1,2)=G^{-1}(1,2)+\Sigma^{e}(1,2),
\end{equation}
one can rewrite Eq.~(\ref{bse}) as follows:
\begin{equation}
R={\tilde{R}}^{0}-i\tilde{R}^{0}WR, \label{bse1}%
\end{equation}
with ${\tilde{R}}^{0}(14,23)={\tilde{G}}(1,3){\tilde{G}}(4,2)$ and $W$ is a
new interaction of the form
\begin{equation}
W=\tilde{V}+W^{e} \label{wampl}%
\end{equation}
where
\begin{equation}
W^{e}(14,23)=V^{e}(14,23)+i\Sigma^{e}(1,3){\tilde{G}}^{-1}(4,2)+i{\tilde{G}%
}^{-1}(1,3)\Sigma^{e}(4,2)-i\Sigma^{e}(1,3)\Sigma^{e}(4,2). \label{wampl1}%
\end{equation}
Since the mean field Green's function ${\tilde{G}}$ in
Eq.~(\ref{G-tilde}) is known one has a more convenient starting
point for the solution of the Bethe-Salpeter equation (\ref{bse1})
than with the unknown exact single-particle Green's function, but
the effective interaction $W$ in Eq.~(\ref{wampl}) becomes more
complicated. The graphical representation of the Eq.~(\ref{bse1}) is
shown in Fig.~\ref{f1}.

In addition to the static interaction $\tilde{V}$ the effective
interaction $W$ contains diagrams with energy dependent
self-energies and an energy dependent \textit{induced interaction},
where a phonon is exchanged between the particle and the hole. Here
and hereinafter we omit the term $i\Sigma ^{e}(3,1)\Sigma^{e}(2,4)$
in the Eq.~(\ref{wampl1}) by the following reasons. This term
compensates the multiple counting of the particle-phonon coupling
arising from the two previous terms in the Eq.~({\ref{wampl1}}).
However, this multiple counting does not take place within time
blocking approximation (see below) if the backward-going propagators
are not taken into account. Components containing the backward-going
propagators within $1p1h\otimes$phonon configurations require a
special consideration. They have been analyzed in great detail in
the Refs. \cite{Tse.89,KTT.97,KST.04}. In the present work they are
fully neglected and therefore the term
$i\Sigma^{e}(3,1)\Sigma^{e}(2,4)$ has also to be omitted. However,
we have to emphasize, that we only neglect ground state correlations
(backward-going diagrams) caused by the particle-phonon coupling.
All the RPA ground state correlations are taken into account,
because it is well known that they play a central role for the
conservation of currents and sum rules. We consider that this is a
reasonable approximation which is applied and discussed also in many
non-relativistic models (see e.g. Refs.
\cite{BBBD.79,BB.81,CBGBB.92,CB.01,SBC.04,Tse.89,KTT.97,KST.04,Tse.05,LT.05}
and references therein).

Considering a Fourier transformation of the Bethe-Salpeter equation
(\ref{bse1}) in time
\begin{align}
R_{k_{1}k_{4},k_{2}k_{3}}(\omega,\varepsilon)  &  ={\tilde{G}}_{k_{1}k_{3}%
}(\varepsilon+\omega){\tilde{G}}_{k_{4}k_{2}}(\varepsilon)\nonumber\\
&  +\sum\limits_{k_{5}k_{6}k_{7}k_{8}}{\tilde{G}}_{k_{1}k_{5}}(\varepsilon
+\omega){\tilde{G}}_{k_{6}k_{2}}(\varepsilon)\int\limits_{-\infty}^{\infty
}\frac{d\varepsilon^{\prime}}{2\pi i}W_{k_{5}k_{8},k_{6}k_{7}}(\omega
,\varepsilon,\varepsilon^{\prime})R_{k_{7}k_{4},k_{8}k_{3}}(\omega
,\varepsilon^{\prime}), \label{bsee}%
\end{align}
one finds that both the solution of this equation $R$ and its kernel
$W$ are singular with respect to the energy variables. Another
difficulty arises because Eq.~(\ref{bse1}) contains integrations
over all time points of the intermediate states. This means that
many configurations which are actually more complex than
$1p1h\otimes$phonon are contained in the exact response function. In
the Ref.~\cite{Tse.89} a special time-projection technique was
introduced to block the $ph$-propagation through these complex
intermediate states. It has been shown that for this type of
response it is possible to reduce the integral equation (\ref{bsee})
to a relatively simple algebraic equation. Obviously, this method
can be applied straightforwardly in our case too.

Starting from the Bethe-Salpeter equation (\ref{bse1}) we divide the problem
to find the exact response function into two parts. First we calculate the
\textit{correlated propagator }$R^{e}$ which describes the propagation under
the influence of the interaction $W^{e}$
\begin{equation}
R^{e}={\tilde{R}}^{0}-i{\tilde{R}}^{0}W^{e}R^{e}. \label{RE}%
\end{equation}
It contains all the effects of particle-phonon coupling and all the
singularities of the integral part of the initial BSE. Second, we have to
solve the remaining equation for the full response function $R$
\begin{equation}
R=R^{e}-iR^{e}\tilde{V}R. \label{respre}%
\end{equation}
Eq.~(\ref{respre}) contains only the static effective interaction
$\tilde{V}$ and can be easily solved if $R^{e}$ is known. Thus, the
main problem is to calculate the correlated propagator $R^{e}$ .

\subsection{Time blocking approximation (TBA) and subtraction method}

In order to calculate the correlated propagator $R^{e}$ we represent this
quantity as an infinite series of graphs which contain mean field
$ph$-propagators alternated with single interaction acts. This can be
expressed by the system of the following equations:
\begin{align}
R^{e}  &  ={\tilde{R}}^{0}-i{\tilde{R}}^{0}{\Gamma}^{e}{\tilde{R}}%
^{0},\label{re}\\
{\Gamma}^{e}  &  =W^{e}-iW^{e}{\tilde{R}}^{0}{\Gamma}^{e}. \label{gamma}%
\end{align}
According to the main idea of the time blocking approximation we
modify the integral part of Eq.~(\ref{gamma}) making use of the
time-projection operator in the form \cite{Tse.89}:
\begin{equation}
\Theta(14,23)=\delta_{\sigma_{k_{1}},-\sigma_{k_{2}}}
\delta_{k_{1}k_{3}}\delta_{k_{2}k_{4}}
\theta(\sigma_{k_{1}}t_{14})\theta(\sigma_{k_{1}}t_{23}),
\label{theta}%
\end{equation}
where $\theta(t)$ is the Heaviside function. This projection
operator is introduced into the mean field propagator
$\tilde{R}^{0}$ in the integral part of the Eq.~(\ref{gamma}) to
order the $ph$-propagation in time and, thus, to separate in time
the acts of the particle-phonon interaction from each other and to
exclude configurations which are more complex than
$1p1h\otimes$phonon.
We replace in Eq.~(\ref{gamma}) the mean field propagator $\tilde{R}^{0}$ by%
\begin{equation}
\tilde{R}^{0}(14,23)\rightarrow\tilde{R}^{0}(14,23)\Theta(14,23)
\end{equation}
and obtain instead of Eq.~(\ref{gamma})
\begin{equation}
{\Gamma}^{e}(14,23)=W^{e}(14,23)+\frac{1}{i}\sum\limits_{5678}W^{e}%
(16,25){\tilde{R}}^{0}(58,67)\Theta(58,67){\Gamma}^{e}(74,83). \label{gamma1}%
\end{equation}
After a Fourier transformation in time we restrict ourselves to the response
function $R_{k_{1}k_{4},k_{2}k_{3}}(\omega)$
\begin{equation}
R_{k_{1}k_{4},k_{2}k_{3}}(\omega)=-i\int\limits_{-\infty}^{\infty}dt_{1}%
dt_{2}dt_{3}dt_{4}\delta(t_{1}-t_{2})\delta(t_{3}-t_{4})\delta(t_{4}%
)e^{i\omega t_{13}}R(14,23), \label{tbaresp}%
\end{equation}
which depends only on one energy variable $\omega$.

The time projection by the operator (\ref{theta}) leads to a
separation of integrations in Eq.~(\ref{bsee}) and we find an
algebraic equation for the function (\ref{tbaresp}):
\begin{equation}
R_{k_{1}k_{4},k_{2}k_{3}}(\omega)=\tilde{R}_{k_{1}k_{4},k_{2}k_{3}}^{0}%
(\omega)+\sum\limits_{k_{5}k_{6}k_{7}k_{8}}\tilde{R}_{k_{1}k_{6},k_{2}k_{5}%
}^{0}(\omega){\bar W}_{k_{5}k_{8},k_{6}k_{7}}(\omega)R_{k_{7}k_{4},k_{8}k_{3}}%
(\omega), \label{respdir}%
\end{equation}
or in shorthand notation
\begin{equation}
R(\omega)=\tilde{R}^{0}(\omega)+\tilde{R}^{0}(\omega){\bar W}(\omega)R(\omega),
\end{equation}
where
\begin{equation}
{\bar W}_{k_{1}k_{4},k_{2}k_{3}}(\omega)=\tilde{V}_{k_{1}k_{4},k_{2}k_{3}}%
+\Phi_{k_{1}k_{4},k_{2}k_{3}}(\omega) \label{W-omega}%
\end{equation}
and
\begin{equation}
{\tilde{R}}_{k_{1}k_{4},k_{2}k_{3}}^{0}(\omega)={\tilde{R}}_{k_{1}k_{2}%
}(\omega)\delta_{k_{1}k_{3}}\delta_{k_{2}k_{4}}. \label{R0}%
\end{equation}
${\tilde{R}}_{k_{1}k_{2}}(\omega)$ is the mean field propagator:
\begin{align}
{\tilde{R}}_{ph}(\omega)  &  =-\frac{1}{\varepsilon_{ph}-\omega}%
,\ \ \ \ {\tilde{R}}_{{\alpha}h}(\omega)=-\frac{1}{\varepsilon_{{\alpha}%
h}-\omega},\\
{\tilde{R}}_{hp}(\omega)  &  =-\frac{1}{\varepsilon_{ph}+\omega}%
,\ \ \ \ {\tilde{R}}_{h{\alpha}}(\omega)=-\frac{1}{\varepsilon_{{\alpha}%
h}+\omega},
\end{align}
$\varepsilon_{ph}=\varepsilon_{p}-\varepsilon_{h}$ and $\Phi$ is the
particle-phonon coupling amplitude with the following components:
\begin{align}
\Phi_{ph^{\prime},hp^{\prime}}(\omega)  &  =\sum\limits_{\mu}\Bigl[\delta
_{pp^{\prime}}\sum\limits_{h^{\prime\prime}}\frac{\gamma_{h^{\prime\prime}%
h}^{\mu}\gamma_{h^{\prime\prime}h^{\prime}}^{\mu\ast}}{\omega-\varepsilon
_{p}+\varepsilon_{h^{\prime\prime}}-\Omega^{\mu}}\nonumber\\
&  +\delta_{hh^{\prime}}\Bigl(\sum\limits_{p^{\prime\prime}}\frac
{\gamma_{pp^{\prime\prime}}^{\mu}\gamma_{p^{\prime}p^{\prime\prime}}^{\mu\ast
}}{\omega-\varepsilon_{p^{\prime\prime}}+\varepsilon_{h}-\Omega^{\mu}}%
+\sum\limits_{{\alpha}^{\prime\prime}}\frac{\gamma_{p{\alpha}^{\prime\prime}%
}^{\mu}\gamma_{p^{\prime}{\alpha}^{\prime\prime}}^{\mu\ast}}{\omega
-\varepsilon_{{\alpha}^{\prime\prime}}+\varepsilon_{h}-\Omega^{\mu}%
}\Bigr)\nonumber\\
&  -\Bigl(\frac{\gamma_{pp^{\prime}}^{\mu}\gamma_{hh^{\prime}}^{\mu\ast}%
}{\omega-\varepsilon_{p^{\prime}}+\varepsilon_{h}-\Omega^{\mu}}+\frac
{\gamma_{p^{\prime}p}^{\mu\ast}\gamma_{h^{\prime}h}^{\mu}}{\omega
-\varepsilon_{p}+\varepsilon_{h^{\prime}}-\Omega^{\mu}}\Bigr)\Bigr],
\label{phiph}%
\end{align}%
\begin{align}
\Phi_{\alpha h^{\prime},h\alpha^{\prime}}(\omega)  &  =\sum\limits_{\mu
}\Bigl[\delta_{\alpha\alpha^{\prime}}\sum\limits_{h^{\prime\prime}}%
\frac{\gamma_{h^{\prime\prime}h}^{\mu}\gamma_{h^{\prime\prime}h^{\prime}}%
^{\mu\ast}}{\omega-\varepsilon_{\alpha}+\varepsilon_{h^{\prime\prime}}%
-\Omega^{\mu}}\nonumber\\
&  +\delta_{hh^{\prime}}\Bigl(\sum\limits_{\alpha^{\prime\prime}}\frac
{\gamma_{\alpha\alpha^{\prime\prime}}^{\mu}\gamma_{\alpha^{\prime}%
\alpha^{\prime\prime}}^{\mu\ast}}{\omega-\varepsilon_{\alpha^{\prime\prime}%
}+\varepsilon_{h}-\Omega^{\mu}}+\sum\limits_{p^{\prime\prime}}\frac
{\gamma_{\alpha p^{\prime\prime}}^{\mu}\gamma_{\alpha^{\prime}p^{\prime\prime
}}^{\mu\ast}}{\omega-\varepsilon_{p^{\prime\prime}}+\varepsilon_{h}%
-\Omega^{\mu}}\Bigr)\nonumber\\
&  -\Bigl(\frac{\gamma_{\alpha\alpha^{\prime}}^{\mu}\gamma_{hh^{\prime}}%
^{\mu\ast}}{\omega-\varepsilon_{\alpha^{\prime}}+\varepsilon_{h}-\Omega^{\mu}%
}+\frac{\gamma_{\alpha^{\prime}\alpha}^{\mu\ast}\gamma_{h^{\prime}h}^{\mu}%
}{\omega-\varepsilon_{\alpha}+\varepsilon_{h^{\prime}}-\Omega^{\mu}%
}\Bigr)\Bigr]. \label{phiah}%
\end{align}
Indices $p,\alpha$ and $h$ traditionally denote the particle,
antiparticle and hole types of the Dirac states, respectively. There
are, in principle, also non-zero amplitudes of the types
$\Phi_{ph^{\prime},h{\alpha}}$, $\Phi_{{\alpha}h^{\prime},hp}$ which
cause transitions of particle-hole ($ph$) to antiparticle-hole
(${\alpha}h$) pairs. However, in the present work we neglect them
because as it has been investigated in Ref.~\cite{LR.06} the effect
of this kind of terms on the self-energy is very small whereas they
require a lot of numerical effort. The reason is that these
components as well as the component (\ref{phiah}) contain large
energy denominators. As it was
already mentioned, the amplitudes of the types $\Phi_{pp^{\prime},hh^{\prime}%
}$ and $\Phi_{hh^{\prime},pp^{\prime}}$ are also disregarded within our
approximation. Therefore ground state correlations are taken into account only
on the RPA level due to the presence of the $\tilde{V}_{pp^{\prime}%
,hh^{\prime}}$, $\tilde{V}_{hh^{\prime},pp^{\prime}}$ terms of the
static interaction in the Eq.~(\ref{respdir}). By definition, the
propagator $R(\omega)$ in Eq.~(\ref{respdir}) contains only
configurations which are not more complex than $1p1h\otimes$phonon.

An important correction has to be done in the Eq.~(\ref{respdir}).
Being adjusted to experimental data the RMF ground state contains
effectively many correlations and, in particular, also admixtures of
phonons. Therefore in our approach, as well as in other approaches
beyond RPA where more complex configurations are taken into account
explicitly, this admixtures would lead to double counting of
correlations. In the present method, all the correlations entering
through the admixture of phonons are taken care of by an additional
interaction term: $\ \Phi(\omega)$. Part of this interaction is
therefore already contained in the static mean field interaction
$\tilde{V}$. \ Since the parameter of the density functional and as
a consequence the effective interaction $\tilde{V}$ is adjusted to
experimental ground state properties at the energy $\omega=0$, this
part of the interaction $\Phi (\omega)$, which is already contained
in $\tilde{V}$ is given by $\Phi(0)$. In order to avoid double
counting of correlations a subtraction procedure has been developed
in the Ref.~\cite{Tse.05} were this part is removed. As a
consequence one has to replace the effective interaction
(\ref{W-omega}) of the response equation (\ref{respdir}) by the
function $\delta\Phi$:
\begin{equation}
\Phi(\omega)\rightarrow\delta\Phi(\omega)=\Phi(\omega)-\Phi(0)\mathbf{.}
\label{subtraction-procedure}%
\end{equation}
The physical meaning of this subtraction is clear: the average value
of the particle-vibration coupling amplitude $\Phi$ at the ground
state is supposed to be contained already in the residual
interaction $\tilde{V}$, therefore we should take into account only
the additional energy dependence, i.e.
$\delta\Phi(\omega)=\Phi(\omega)-\Phi(0)$ on top of the effective
interaction $\tilde{V}$. Instead of Eq.~(\ref{respdir}) we finally
solve the following response equation
\begin{equation}
R=\tilde{R}^{0}+\tilde{R}^{0}[\tilde{V}+\delta\Phi]R. \label{resp-subtr}%
\end{equation}

\subsection{Strength function and transition densities}

To describe the observed spectrum of the excited nucleus in a weak external
field $P$, as for instance a dipole field, one needs to calculate the strength
function:
\begin{equation}
S(E)=-\frac{1}{\pi}\lim\limits_{\Delta\rightarrow+0}Im\ \Pi_{PP}(E+i\Delta),
\label{strf}%
\end{equation}
expressed through the polarizability $\Pi_{PP}$ defined as
\begin{equation}
\Pi_{PP}(\omega)=P^{\dag}R(\omega)P:=\sum\limits_{k_{1}k_{2}k_{3}k_{4}%
}P_{k_{1}k_{2}}^{\ast}R_{k_{1}k_{4},k_{2}k_{3}}(\omega)P_{k_{3}k_{4}}.
\label{polarization}%
\end{equation}
The imaginary part $\Delta$ of the energy variable is introduced for
convenience in order to obtain a more smoothed envelope of the spectrum. This
parameter has the meaning of an additional artificial width for each
excitation. This width emulates effectively contributions from configurations
which are not taken into account explicitly in our approach.

In order to calculate the strength function it is convenient to
convolute Eq.~(\ref{resp-subtr}) with an external field operator and
introduce the density matrix variation $\delta\rho$ in the external
field $P$:
\begin{align}
\delta\rho_{k_{1}k_{2}}(\omega)  &  =\sum\limits_{k_{3}k_{4}}R_{k_{1}%
k_{4},k_{2}k_{3}}(\omega)P_{k_{3}k_{4}},\label{DrhoP}\\
{\delta\rho}_{k_{1}k_{2}}^{0}(\omega)  &  =\sum\limits_{k_{3}k_{4}}{\tilde{R}%
}_{k_{1}k_{4},k_{2}k_{3}}^{0}(\omega)P_{k_{3}k_{4}}, \label{DrhoP0}%
\end{align}
Using Eq.~(\ref{resp-subtr}) we find that $\delta\rho(\omega)$ obeys
the equation
\begin{equation}
\delta\rho(\omega)={\delta\rho}^{0}(\omega)+\tilde{R}^{0}(\omega
)\Bigl(\tilde{V}+\delta\Phi(\omega)\Bigr){\delta\rho}(\omega), \label{drho1}%
\end{equation}
and the strength function is expressed as%
\begin{equation}
S(E)=-\frac{1}{\pi}\lim\limits_{\Delta\rightarrow+0}Im\,
\text{Tr[}P^{\dag}\delta
\rho(E+i\Delta)\text{]}. \label{strf1}%
\end{equation}

The Bethe-Salpeter Eq.~(\ref{resp-subtr}) gives us also the
possibility to
calculate the transition density%
\begin{equation}
\rho_{k_{1}k_{2}}^{\nu}=\langle0|\psi_{k_{2}}^{\dagger}\psi_{k_{_{1}}}%
|\nu\rangle
\end{equation}
for the excited state $|\nu\rangle$ at the energy $\Omega^{\nu}$. In the
vicinity of $\Omega^{\nu}$ the response function has a simple pole structure
\begin{equation}
R_{k_{1}k_{4},k_{2}k_{3}}^{\nu}(\omega)\approx \frac{\rho_{k_{1}k_{2}}^{\nu
}\rho_{k_{3}k_{4}}^{{\nu}\ast}}{\omega-\Omega^{\nu}}\;. \label{pole-structure}%
\end{equation}
which leads to
\begin{equation}
\rho_{k_{1}k_{2}}^{\nu}=\lim\limits_{\Delta\rightarrow+0}\sqrt{\frac{\Delta
}{\pi S(\Omega^{\nu})}}\ Im\ \delta\rho_{k_{1}k_{2}}(\Omega^{\nu}+i\Delta),
\label{ampl}%
\end{equation}
In order to determine the norm of the transition densities, it is convenient
to rewrite Eq.~(\ref{resp-subtr}) in the following form:%
\begin{equation}
\Bigl(({\tilde{R}}^{0}{)}^{-1}(\omega)-\tilde{V}-\delta\Phi(\omega
)\Bigr)R(\omega)=1 \label{bsem1}%
\end{equation}
Using Eq.~(\ref{pole-structure}) and taking the derivative with
respect to
$\omega$, we obtain the generalized normalization condition:%
\begin{equation}
\rho_{{}}^{{\nu}\ast}\Bigl[N-\frac{d\Phi}{d\omega}\mid_{\omega=\Omega^{\nu}%
}\Bigr]\rho^{\nu}=1\;, \label{norm}%
\end{equation}
with the RPA norm
\[
N_{k_{1}k_{4},k_{2}k_{3}}=\sigma_{k_{1}}
\delta_{\sigma_{k_{1}},-\sigma_{k_{2}}}
\delta_{k_{1}k_{3}}\delta_{k_{2}k_{4}}.%
\]
In the limiting case of an energy-independent interaction, i.e. if one
neglects $\Phi(\omega)$, this reduces to the usual RPA normalization:
\begin{equation}
\sum\limits_{ph}\Bigl(|\rho_{ph}^{\nu}|^{2}-|\rho_{hp}^{\nu}|^{2}\Bigr)=1\;.
\end{equation}
In the particle-vibration coupling model the quantity
%%%derivative $d\Phi/d\omega$
$N \frac{d}{d\omega} \Phi$
in the Eq.~(\ref{norm}) is a non-positively definite
matrix. Therefore, all the eigenvalues of the operator
$[1 - N\,(\frac{d}{d\omega} \Phi)\mid_{\omega=\Omega^{\nu}}]^{-1}$
are less than or equal to 1,
in analogy to the spectroscopic factor of a
single-particle state. For this reason the sum
\begin{equation}
\sum\limits_{ph}\Bigl(|\rho_{ph}^{\nu}|^{2}-|\rho_{hp}^{\nu}|^{2}\Bigr)
\end{equation}
is practically always less than 1.
The reduction of the norm is caused by the spreading of
the strength over $1p1h\otimes$phonon configurations.

\section{Numerical details, results, and discussion}

\label{results}
\subsection{General scheme of the calculations}

The method developed in the last section is applied to a quantitative
description of isoscalar monopole and isovector dipole giant resonances in the
even-even spherical nuclei $^{208}$Pb and $^{132}$Sn. Our calculations are
based on the energy functional with the non-linear parameter set NL3
\cite{NL3}. The scheme consists of three main parts:

i) The Dirac equation for single nucleons together with the Klein-Gordon
equations for meson fields are solved simultaneously in a self-consistent way
to obtain the single-particle basis (Dirac basis).

ii) The RRPA equations \cite{RMG.01} are solved with the static
interaction $\tilde{V}$ of Eq.~(\ref{static-interaction}) to
determine the low-lying collective vibrations (phonons). These two
sets of particles (holes) and phonons form the multitude of
$1p1h\otimes$phonon configurations which enter the particle-phonon
coupling amplitude $\delta\Phi(\omega)$.

iii) Eq.~(\ref{drho1}) for the density matrix variation $\delta\rho$
in the external field $P$ is solved using this additional amplitude
in the effective interaction $\tilde{V}+\delta\Phi(\omega)$.
Eq.~(\ref{strf1}) finally allows to calculate the strength function
corresponding to the operator $P$. It is found that the energy
dependent term $\delta\Phi(\omega)$, which describes the change of
the effective interaction due to the energy dependence of the
particle-vibration coupling, provides a considerable enrichment of
the calculated spectrum as compared to the pure RRPA.

\subsection{Choice of representation and basic approximations}

%%%Eq.~(\ref{drho1}) for the response function can be solved in various
Eq.~(\ref{drho1}) for $\delta\rho$ can be solved in various
representations. In Dirac space its dimension is the number of
$ph$-pairs. In relativistic nuclear calculations it is often
important to take into account the contributions of the Dirac sea
and then the total number of $ph$ and $\alpha h$ pairs entering the
Eq.~(\ref{drho1}) increases considerably with the nuclear mass
number because of the high level density. As it was investigated in
a series of RRPA calculations \cite{RMG.01,MWG.02}, the completeness
of the $ph$ ($\alpha h$) basis is very important for calculations of
giant resonance characteristics as well as for current conservation
and a proper treatment of symmetries, in particular, the dipole
spurious state originating from the violation of translation
symmetry on the mean field level. On the other side, the use of a
large basis requires a considerable numerical effort and, therefore
it is reasonable to solve the Eq.~(\ref{drho1}) in a different more
appropriate representation.

Two facts simplify our practical calculations:

i) Due to the pole structure of the $\delta\Phi$ amplitude it turns out that
the effects of particle-phonon coupling are only important in the vicinity of
the Fermi surface and therefore one can restrict the number of $ph$%
-configurations in this case much more than in the case of the static
interaction $\tilde{V}.$

ii) The effective interaction ${\tilde{V}}$, which is based on the exchange of
mesons, contains only direct terms and no exchange terms. Therefore it can be
written as a sum of separable interactions.

It order to exploit these two effects we solve the response equation for a
fixed value of the energy $\omega$ in two steps. First we calculate the
\textit{correlated propagator }$R^{e}(\omega)$ which describes the propagation
under the influence of the interaction $\delta\Phi(\omega)$
\begin{equation}
R^{e}={\tilde{R}}^{0}+{\tilde{R}}^{0}\delta\Phi R^{e}. \label{RE1}%
\end{equation}
It contains all the effects of particle-phonon coupling and all the
singularities of the integral part of the initial BSE. Since
$\delta\Phi$ is not separable Eq.~(\ref{RE1}) has to be solved in
Dirac space. This requires in principle for each value of $\omega$
the inversion of a very large matrix. However the numerical effort
can be reduced considerably by taking into account that the effects
of the particle-phonon coupling are only important in the vicinity
of the Fermi surface. Therefore the summation in the Eq.~(\ref{RE1})
is performed only among the $ph$-pairs with
$\varepsilon_{ph}\leq\varepsilon_{win}$. Consequently, the
correlated propagator $R^{e}$ differs from the mean field propagator
$\tilde{R}^{0}$ only within this window. This approximation has been
investigated numerically by explicit calculations with different
window sizes as it is discussed below. It is important to emphasize
that the number of $ph$- and $\alpha h$-configurations taken into
account on the RRPA level has to be considerably larger in order to
obtain convergence in the centroid positions of giant resonances as
well as to find the dipole spurious state close to zero energy.

In the second step, we have to solve the remaining equation for the full
response function $R(\omega)$:
\begin{equation}
R=R^{e}+R^{e}\tilde{V}R. \label{RVR}%
\end{equation}
%%%Instead of
In contrast to
the Eq.~(\ref{resp-subtr}) this equation contains only
the static effective interaction.

Now we exploit the fact that the one-boson exchange (OBE)
interaction discussed in Appendix~\ref{choice} is separable in
momentum space. It can be expressed as%
\begin{equation}
\tilde{V}_{k_{1}k_{4},k_{2}k_{3}}=\sum\limits_{c}d_{c}Q_{k_{1}k_{2}}%
^{(c)}Q_{k_{3}k_{4}}^{(c)\ast}%
%%%\tilde{V}_{k_{1}k_{4},k_{2}k_{3}}=\sum\limits_{c}d_{c}Q_{k_{2}k_{1}}%
%%%^{(c)\ast}Q_{k_{4}k_{3}}^{(c)}%
\end{equation}
where the channel index $c=\left(  q,m\right)  $ is given by the
momentum $q$ transferred in the exchange process of the
corresponding meson labeled by the index $m$. \ The parameters
$d_{c}$ are given by the meson propagators. We now can use the well
known techniques of the response formalism with separable
interactions (see, for instance, Ref.~\cite{RS.80}). We define%
\begin{equation}
R_{cc^{\prime}}(\omega)=\sum\limits_{k_{1}k_{2}k_{3}k_{4}}Q_{k_{1}k_{2}}%
^{(c)\ast}R_{k_{1}k_{4},k_{2}k_{3}}(\omega)Q_{k_{3}k_{4}}^{(c^{\prime})} \label{RCC}%
\end{equation}
and find from Eq.~(\ref{RVR})%
\begin{equation}
R_{cc^{\prime}}^{{}}=R_{cc^{\prime}}^{e}+\sum\limits_{c^{\prime\prime}%
}R_{cc^{\prime\prime}}^{e}d_{c^{\prime\prime}}^{{}}R_{c^{\prime\prime
}c^{\prime}}^{{}}=R_{cc^{\prime}}^{e}+(R^{e}dR)_{cc^{\prime}}.
\end{equation}
This equation can be solved by matrix inversion%
\begin{equation}
R=\Bigl(1-R^{e}d\Bigr)^{-1}R^{e}. \label{bse3}%
\end{equation}
Finally we transform Eq.~(\ref{drho1}) to the momentum space, which
is equivalent to using the external field as an additional channel
$c=p$, and obtain from Eq.~(\ref{RCC}) the full response
\begin{equation}
\delta\rho_{c}=\delta\rho_{c}^{e}+\sum\limits_{c^{\prime}}R_{cc^{\prime}}%
^{e}d_{c^{\prime}}^{{}}\delta\rho_{c^{\prime}} \label{drho4}%
\end{equation}
and the polarizability (\ref{polarization}):%
\begin{equation}
\Pi_{PP}=P^{\dag}RP=P^{\dag}R^{e}P+\sum\limits_{c}\delta\rho_{c}^{e\ast}d_{c}^{{}%
}\delta\rho_{c}^{{}}=R_{p^{{}}p}^{e}+\sum\limits_{cc^{\prime}}\delta\rho_{c}^{e\ast}%
d_{c}^{{}}\Bigl(1-R^{e}d\Bigr)_{cc^{\prime}}^{-1}\delta\rho_{c^{\prime}}^{e}.%
\end{equation}
The rank of vectors and matrices entering the Eq.~(\ref{drho4}) is
determined by the number of mesh-points in $q$-space and the number
of meson channels. In particular it does not depend on the mass
number of the nucleus. In realistic calculations the rank of the
Eq.~(\ref{bse3}) in the momentum space is around 500 and, obviously,
the numerical effort does not depend considerably on the total
dimension of $ph$ and $\alpha h$ subspaces. However, if one stays in
the Dirac basis the latter dimension is exactly the rank of arrays
in the Eq.~(\ref{respdir}). One should bear in mind that in practice
both subspaces ($ph$ and $\alpha h$) are truncated at some energy
differences $\varepsilon_{ph}$ and $\varepsilon_{\alpha h}$ which
are large enough so that a further increase of these values does not
influence the results. In light nuclei with relatively small level
density the total dimension of $ph$ and $\alpha h$ subspaces is
comparable and could be even smaller than the rank of the
Eq.~(\ref{drho4}) and, therefore, working in the Dirac basis is more
preferable. For heavy nuclei the dimension of the Dirac basis
becomes huge and, therefore, using of the momentum space is more
justified.

\subsection{Numerical details}

To ensure numerical properness of our codes the response equation
for $\delta\rho$ has been solved both in momentum space
(\ref{drho4}) and in Dirac space (\ref{drho1}) and identical results
have been obtained. Both Fermi and Dirac subspaces were truncated at
energies far away from the Fermi surface: in the present work as
well as in the Ref.~\cite{LR.06} we fix the limits
$\varepsilon_{ph}<100$ MeV and $\varepsilon_{\alpha h}>-1800$ MeV
with respect to the positive continuum. A small artificial width of
200 keV was introduced as an imaginary part of the energy variable
$\omega$ to have a smooth envelope of the calculated curves. The
energies and amplitudes of the most collective phonon modes with
spin and parity 2$^{+}$, 3$^{-}$, 4$^{+}$, 5$^{-}$, 6$^{+}$ have
been calculated with the same restrictions and selected using the
same criterion as in the Ref.~\cite{LR.06} and in many other
non-relativistic investigations in this context. Only the phonons
with energies below the neutron separation energy enter the phonon
space since the contributions of the higher-lying modes are found to
be small. Test calculations in the framework of the approach
\cite{Tse.05,LT.05}
without the restriction of the phonon space by
the energy resulted in the small deviation of the strength function as well as
the change of the mean energies and widths of the resonances comparable
with the smearing parameter (imaginary part of the energy variable)
used in the calculations. This is the natural
result because the physical sense of this parameter is to emulate
contributions of the configurations which are not taken into account
explicitly.

Because of the pole structure of the particle-phonon coupling amplitude
(\ref{phiph}) and (\ref{phiah}) its contributions to the final result decrease
considerably when we go away from the Fermi surface. Therefore this coupling
has been taken into account only within the $ph$-energy window $\varepsilon
_{ph}\leq$ 30 MeV around the Fermi surface. It has been checked that
further increase of this window does not influence considerably the results.

It is important to note that although a large number of configurations of the
$1p1h\otimes$phonon type are taken into account explicitly in our approach,
nevertheless we stay in the same $ph$ ($\alpha h$) space as in the RRPA,
therefore the problem of completeness of the phonon basis does not arise and,
so, the phonon subspace can be truncated in the above mentioned way. Another
essential point is, that on all three stages of our calculations the same
relativistic nucleon-nucleon interaction $\tilde{V}$
\ (\ref{static-interaction}) has been employed. The vertices $\gamma
_{k_{1}k_{2}}^{\mu}$ (\ref{phonon}) entering the term $\delta\Phi(\omega)$ are
calculated with the same force. Therefore no further parameters are needed.
Our calculational scheme is fully consistent.

The subtraction procedure (\ref{subtraction-procedure}) developed in
the Ref.~\cite{Tse.05} for the self-consistent scheme has been
incorporated in our approach. As it was mentioned above, this
procedure removes the static contribution of the particle-phonon
coupling from the $ph$-interaction. It takes into account only the
additional energy dependence introduced by the dynamics of the
system. It has been found in the present calculations as well as in
the calculations of the Ref.~\cite{LT.05} that within the relatively
large energy interval (0 - 30 MeV) the subtraction procedure
provides a rather small increase of the mean energy of the giant dipole
resonance (0.8 MeV for lead region) and gives rise to
the change by a few percents in the
sum rule. This procedure
%does not change
restores the response at zero energy
and therefore it does not disturb the symmetry properties of the
RRPA calculations. The zero energy modes connected with the
spontaneous symmetry breaking in the mean field solutions, as for
instance the translational mode in the dipole case, remain at
exactly the same position after the inclusion of the
particle-vibration coupling. In practice, however, because of the
limited number of oscillator shells in our calculations this state
is found already in RRPA without particle-vibration coupling at a
few hundreds keV above zero. In cases, where the results depend
strongly on a proper separation of this spurious state, as for
instance for investigations of the pygmy dipole resonance in neutron
rich systems, we have to include a large number of\ $ph$
-configurations in the RRPA solution.

\begin{figure}[ptb]
\begin{center}
\includegraphics*[scale=1.5]{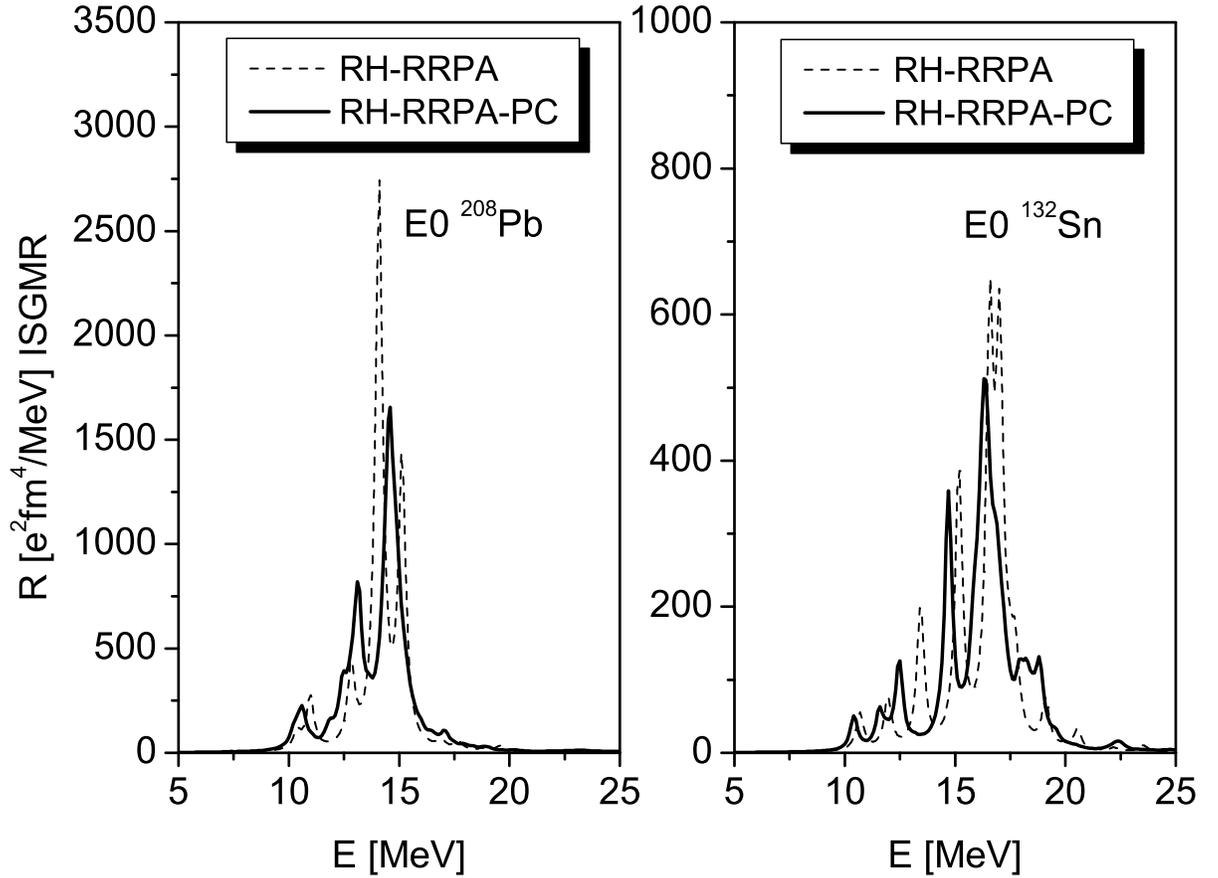}
\end{center}
\caption{Isoscalar monopole resonance in $^{208}$Pb and $^{132}$Sn
obtained within two approaches: RRPA (dashed line) and RRPA with
particle-phonon coupling RRPA-PC (solid line). Both computations
have been performed with relativistic Hartree (RH) mean field and
employ NL3 parameter
set for RMF forces.}%
\label{f2}%
\end{figure}

\begin{table}[ptb]
\caption{Lorentz fit parameters of isoscalar E0 strength function in $^{208}%
$Pb and $^{132}$Sn calculated within RRPA and RRPA extended by the
particle-phonon coupling model (RRPA-PC) as compared to experimental
data. The
fit has been carried out in the interval from $B_{n}$ to roughly 20 MeV.}%
\label{tab1}
\begin{center}
\vspace{3mm} \tabcolsep=2.15em \renewcommand{\arraystretch}{1.1}%
\begin{tabular}
[c]{cccc}\hline\hline &  & $<$E$>$ (MeV) & $\Gamma$ (MeV)\\\hline
& RRPA & 14.16 & 1.71\\
$^{208}$Pb & RRPA-PC & 14.05 & 2.36\\
& Exp. \cite{SY.93} & 13.73(20) & 2.58(20)\\\hline
& RRPA & 16.10 & 2.63\\
$^{132}$Sn & RRPA-PC & 16.01 & 3.09\\\hline\hline
\end{tabular}
\end{center}
\end{table}

\subsection{Isoscalar monopole and isovector dipole resonances in $^{208}$Pb and $^{132}$Sn}

The calculated strength functions for the isoscalar monopole
resonance in $^{208}$Pb and $^{132}$Sn are shown in Fig.~\ref{f2}.
The fragmentation of the resonance caused by the particle-phonon
coupling is clearly demonstrated although the spreading width of the
monopole resonance is not large because of a strong cancellation
between the self-energy diagrams and diagrams with the phonon
exchange (see Fig.~\ref{f1}). This fact has also been discussed in
detail in Refs. \cite{BBBD.79,BB.81} and it is not disturbed by the
subtraction procedure (\ref{subtraction-procedure}) because this
cancellation takes place as well in $\Phi(\omega)$ as in $\Phi(0)$.

In order to compare the spreading of the theoretical strength
distributions with experimental data we deduce mean energies
$\langle E\rangle$ and widths parameters $\Gamma$ by fitting our
theoretical strength distribution in a certain energy interval to a
Lorentz curve in the same way as it has been done in the
experimental investigations. The mean energies and the widths
parameters obtained in this way are shown in the Table~\ref{tab1}.
As experimental data we display the numbers adopted in the
Ref.~\cite{SY.93} from the evaluation of a series of data obtained
in different experiments for the isoscalar monopole resonance in
$^{208}$Pb.

\begin{figure}[ptb]
\begin{center}
\includegraphics*[scale=1.5]{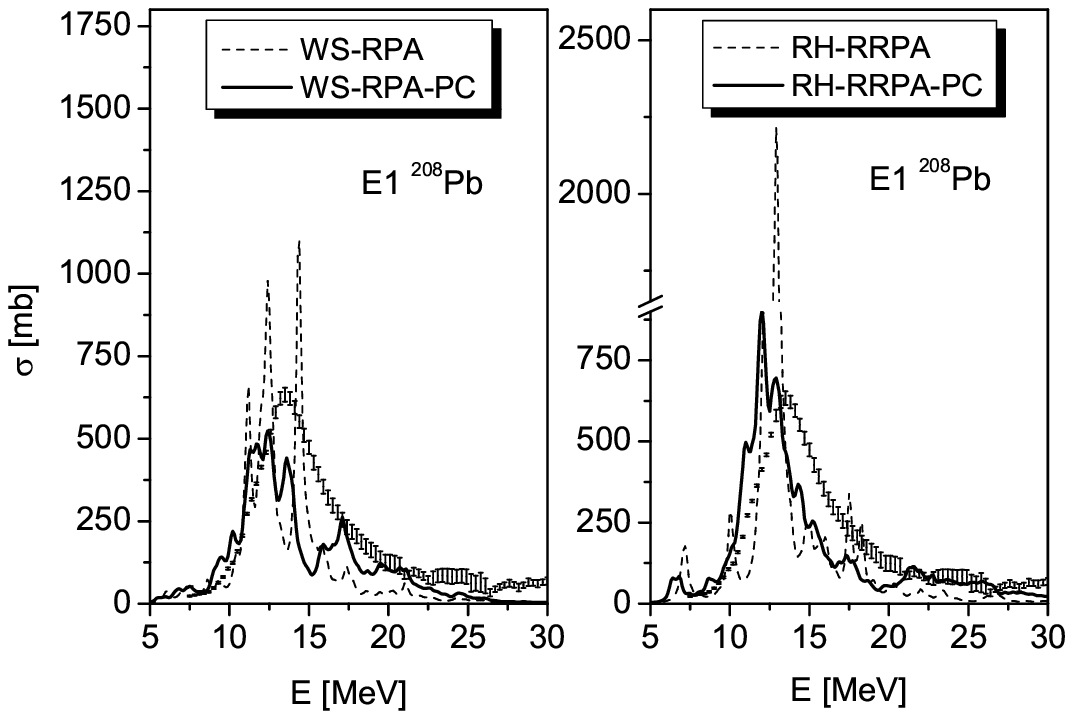}
\end{center}
\caption{The isovector E1 resonance in $^{208}$Pb. Details are given
in the text.}%
\label{f3}%
\end{figure}
\begin{figure}[ptb]
\begin{center}
\includegraphics*[scale=1.5]{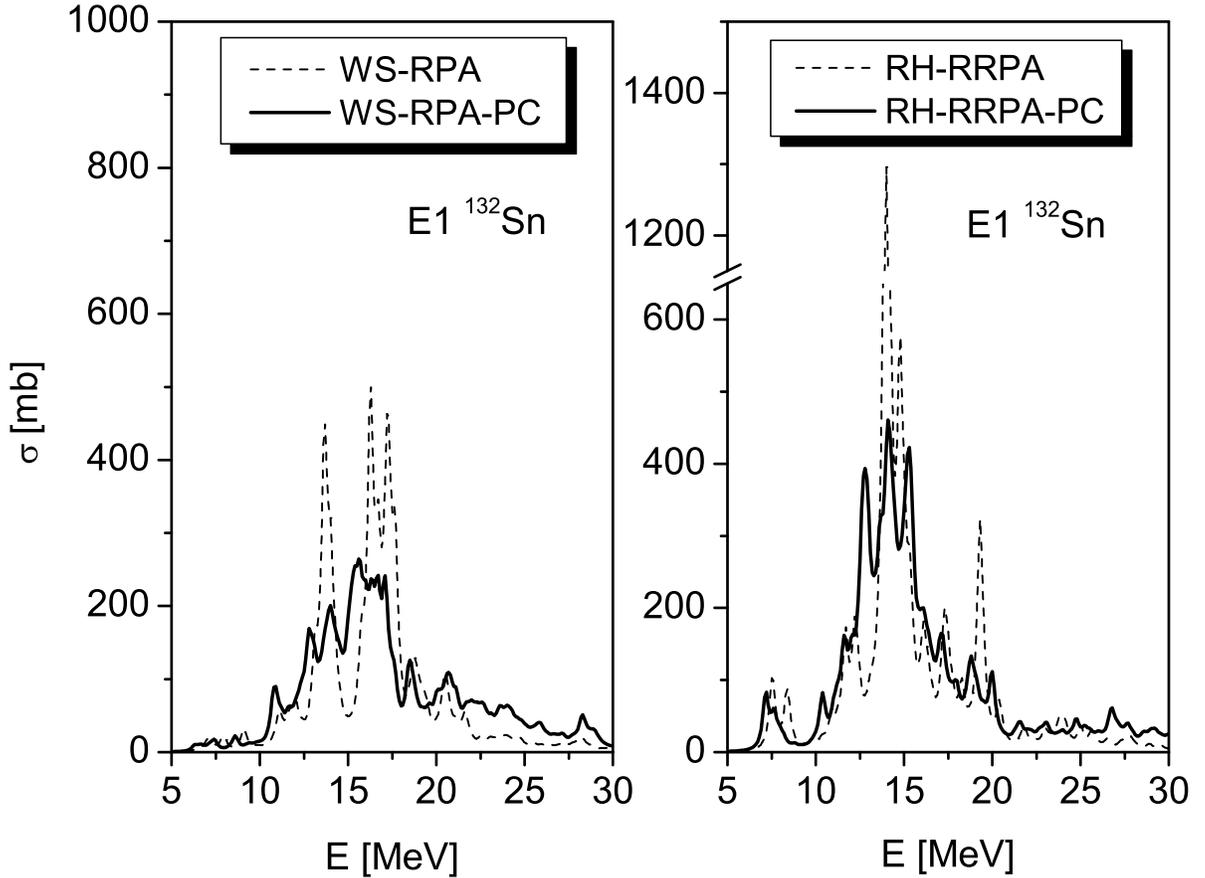}
\end{center}
\caption{The same as in Fig.~\ref{f2} but for $^{132}$Sn.}%
\label{f4}%
\end{figure}

\begin{table}[ptb]
\caption{Lorentz fit parameters in the two energy intervals: $B_{n}-25$ MeV
and $0-30$ MeV, for the E1 photoabsorption cross sections in $^{208}$Pb and
$^{132}$Sn, calculated with the RRPA, and with the RRPA extended to include
the particle-phonon coupling (RRPA-PC), compared to data.}%
\label{tab2}
\begin{center}
%\vspace{6mm}
\tabcolsep=1.1em \renewcommand{\arraystretch}{1.1}%
\begin{tabular}
[c]{cccccccc}\hline\hline
&  & \multicolumn{3}{c}{$B_{n}$ - 25 MeV} & \multicolumn{3}{c}{0 - 30 MeV}\\
&  & $<$E$>$ & $\Gamma$ & EWSR & $<$E$>$ & $\Gamma$ & EWSR\\
&  & (MeV) & (MeV) & (\%) & (MeV) & (MeV) & (\%)\\\hline
& RRPA & 13.1 & 2.4 & 121 & 12.9 & 2.0 & 128\\
$^{208}$Pb & RRPA-PC & 12.9 & 4.3 & 119 & 13.2 & 3.0 & 128\\
& Exp. \cite{ripl} & 13.4 & 4.1 & 117 &  &  & 125(8)\\\hline
& RRPA & 14.7 & 3.3 & 116 & 14.5 & 2.6 & 126\\
$^{132}$Sn & RRPA-PC & 14.4 & 4.0 & 112 & 14.6 & 3.2 & 126\\
& Exp. \cite{Adr.05} & 16.1(7) & 4.7(2.1) & 125(32) &  &  & \\\hline\hline
\end{tabular}
\end{center}
\end{table}

The calculated photoabsorption cross sections
\begin{equation}
\sigma_{E1}(E)={\frac{{16\pi^{3}e^{2}}}{{9\hbar c}}}E~S_{E1}(E)
\end{equation}
for the isovector dipole resonance in $^{208}$Pb and $^{132}$Sn are
given in the Figs. \ref{f3} and \ref{f4} respectively. The left
panels show the results obtained within the non-relativistic
approach with Woods-Saxon (WS) single-particle input and
Landau-Migdal (LM) forces described in Ref.~\cite{LT.05}. They are
compared in the right panel with the relativistic fully consistent
theory developed in the present work. The underlying Lagrangian has
the parameter set NL3~\cite{NL3}. RPA calculations are shown by the
dashed curves, RPA-PC calculations -- by the thick solid curves.
In Fig. \ref{f3} we have displayed experimental data with error bars taken
from Ref. \cite{ripl}. To make
the comparison reasonable calculations within the non-relativistic
framework have been performed with box boundary conditions for the
Schr\"{o}dinger equation in $r$-space which ensures completeness of the
single-particle basis.

The corresponding Lorentz fit parameters in the two energy
intervals: $B_{n}-25$ MeV and $0-30$ MeV ($B_{n}$ is the neutron
separation energy) are included in Table~\ref{tab2} and they are
compared with the data of Ref.~\cite{Adr.05,ripl}. We notice that
the inclusion of particle-phonon coupling in the RRPA calculation
induces a pronounced fragmentation of the photoabsorption cross
sections, and brings the width of the GDR in much better agreement
with the data, both for $^{208}$Pb and $^{132}$Sn.

The fragmentation of the resonance introduced by the particle-phonon
coupling is clearly demonstrated in both cases. Also, one finds more
or less the same level of agreement between theory and experimental
data for these two calculations. In the case of the isovector E1
resonance in $^{132}$Sn this is, however, not so clear because the cross section
and the integral characteristics of the resonance obtained in the experiment
of Ref.~\cite{Adr.05} are given with relatively large error bars.
In $^{208}$Pb our self-consistent relativistic approach reproduces
the shape of the giant dipole resonance much better than
the non-relativistic one although the whole resonance is about 0.5
MeV shifted to lower energies with respect to the experiment. As one can see from the
Fig. \ref{f3} and Table \ref{tab2}, we observe some shift already in the RRPA calculation,
which is determined by the properties of the NL3 forces. Improvement
of the forces, for instance, the use of the density dependent versions \cite{DD-ME1,DD-ME2}
of the RMF should bring the E1 mean energy in better agreement with the data.

However, there is an essential difference between the fully
self-consistent relativistic calculations of the present work and
the non-relativistic approach: in non-relativistic approach
discussed in Ref.~\cite{LT.05} one introduces on all three stages of
the calculation phenomenological parameters, which have to be
adjusted to experimental data: first, the Woods-Saxon parameters as,
for instance, the well depth are varied to obtain single-particle
levels close to the experimental values, second, one of the
parameters of the Landau-Migdal force is adjusted to get phonon
energies at the experimental positions (for each mode) and, third,
another Landau-Migdal force parameter is varied to reproduce the
centroid of the giant resonance. Although the varying of the
parameters is performed in relatively narrow limits, it is necessary
to obtain realistic results. In contrast, in the relativistic fully
consistent approach developed in the present work no adjustment of
additional parameters is made. Of course, the underlying energy
functional has been determined in a phenomenological way by a fit to
experimental ground state properties of characteristic nuclei.
However, it is of universal nature and the same parameters are used
for investigations of many nuclear properties all over the periodic
table. The predictive power of this scheme is therefore much higher
than that of the present semi-phenomenological approach discussed,
for instance, in Ref.~\cite{LT.05}.

The effect of the particle-vibration coupling on the low-lying
dipole strength around the neutron threshold within the presented
approach can be also seen in the Figs. \ref{f3},\ref{f4}. But this
phenomenon requires the detailed investigation and is considered in
the special publication \cite{LRV.06}.

\section{Summary and outlook}

In order to describe the damping of collective excitations in nuclei covariant
density functional theory is extended by particle-vibration coupling in a
consistent way. Starting from a relativistic energy functional $E[\rho]$ the
self-consistent RMF equations are solved and the static part of the nucleon
self-energy $\tilde{\Sigma}=\delta E/\delta \rho$ is found.
In a second step the
low-lying collective phonons are determined in the framework of relativistic
RPA with the effective $ph$-interaction
$\tilde{V}=\delta \tilde{\Sigma}/\delta \rho$.
Using particle-phonon vertices deduced from the same interaction the
self-energy $\tilde{\Sigma}$ is extended by an energy dependent part
$\Sigma^{e}(\ve)$. The Bethe-Salpeter equation derived from the full
self-energy $\Sigma=\tilde{\Sigma}+\Sigma^{e}$ is formulated in the $ph$-basis
of the Dirac eigenstates. Using the time blocking approximation, which allows
the truncation to $1p1h\otimes$phonon$\ $configurations and a subtraction
procedure, which avoids double counting of correlations it is shown that the
resulting effective interaction $\tilde{V}+\delta\Phi(\omega)$ contains
besides the static part $\tilde{V}$ an energy dependent correction term
$\delta\Phi(\omega)$, which takes into account the coupling of the particles
to collective phonons.

This method is applied to the computation of spectroscopic characteristics of
nuclear excited states in a wide energy range up to 30 MeV for spherical
nuclei with closed shells. An equation for the density matrix variation is
formulated and solved in the Dirac space as well in the momentum space. The
particle-phonon coupling amplitudes of collective vibrational modes below the
neutron separation energy have been calculated within the self-consistent RRPA
using the parameter set NL3 for the Lagrangian. The same force has been used
for the static part of the effective $ph$-interaction $\tilde{V}$ and for the
evaluation of the particle-phonon vertices $\gamma_{kk^{\prime}}^{\mu}$.
Therefore a fully consistent description of giant resonances is performed.

Noticeable fragmentation of the isoscalar monopole and isovector dipole giant
resonances in $^{208}$Pb and $^{132}$Sn is obtained due to the
particle-vibration coupling. This leads to a significant spreading width as
compared to simple RRPA calculations. This is in agreement with experimental
data as well as with the results obtained within the non-relativistic
semi-phenomenological approaches of Refs. \cite{LT.05,SBC.04}.

Thus in the present work a description of nuclear many-body dynamics including
complex configurations is realized within an approach which is (i) fully
consistent, (ii) based on relativistic dynamics, (iii) universally valid for
nuclei all over the periodic table, and (iv) based on a modern covariant
density functional, which has been applied with great success for many nuclear
properties all over the periodic table.

So far, this extended density functional theory for complex
configurations has been formulated only for closed shell nuclei. To
expand the field of applications to nuclei with opened shells it is,
of course, necessary to include pairing correlations into the
approach discussed in the present paper. This can be done in the
manner of Ref.~\cite{Tse.05} where pairing correlations and
particle-vibration coupling are taken into account on an equal
footing in a non-relativistic framework using techniques of a
generalized Green's function formalism. Work on this direction is in
progress.

\bigskip

\leftline{\bf ACKNOWLEDGEMENTS}

Helpful discussions with Prof. D. Vretenar are gratefully acknowledged. This
work has been supported in part by the Bundesministerium f\"{u}r Bildung und
Forschung under project 06 MT 246. E. L. acknowledges the support from the
Alexander von Humboldt-Stiftung and the assistance and hospitality provided by
the Physics Department of TU-M\"{u}nchen. V.~T. acknowledges financial support
from the Deutsche Forschungsgemeinschaft under the grant No. 436 RUS
113/806/0-1 and from the Russian Foundation for Basic Research under the grant
No. 05-02-04005-DFG\_a.
\bigskip

\appendix

\section{Non-linear meson potentials}

\label{nls} So far we started from a Lagrangian (\ref{Lagr}) containing the
nucleons as Dirac particles and mesons providing an interaction between them
\cite{Wal.74,SW.86}, as it is the starting point of most of the relativistic
investigations in nuclear physics. It has been recognized, however, very early
\cite{BB.77} that this model does not provide an adequate description of the
surface properties of realistic nuclei. The incompressibility is too high and
the deformations are too small \cite{GRT.90}. Therefore over the years the
relativistic models have been considerably improved by introducing an
effective density dependence, as it is in accordance with the concept of
density functional theory. Boguta and Bodmer proposed already in 1977 to
introduce a non-linear coupling between the $\sigma$-mesons replacing \ the
mass term $\frac{1}{2}m_{\sigma}^{2}\sigma^{2}$ in the Lagrangian (\ref{Lagr})
by a non-linear meson potential \cite{BB.77}%
\begin{equation}
U(\sigma)=\frac{1}{2}m_{\sigma}^{2}\sigma^{2}+\frac{g_{2}}{3}\sigma^{3}%
+\frac{g_{3}}{4}\sigma^{4}\;. \label{NL}%
\end{equation}
This procedure was used because it did not destroy the renormalizability of
the model. In these early days relativistic models of the nucleus were
considered as fully fledged quantum field theories \cite{SW.86} and therefore
non-linear meson couplings provided an elegant extension of the model
containing additional parameters introducing an effective density dependence.

In fact, models with non-linear meson couplings have been used very
extensively in the literature. Very successful parameterizations
have been developed on this basis as, for instance, NL1 \cite{NL1}
and NL3 \cite{NL3}, which contain only non-linear couplings between
the $\sigma$-mesons. Later on one also considered non-linear
couplings between the $\omega$-mesons \cite{Bod.91,TM1}, however
without improving the agreement between theoretical results and
experimental data in finite nuclei. In the present investigations we
consider the parameter set NL3, which has been used with great
success for the description of symmetric nuclear matter and finite
nuclei with closed shells \cite{GRT.90}, deformed nuclei, rotating
nuclei \cite{ARK.00} and for relativistic RPA calculations of giant
resonances and low-lying states \cite{MWG.02}. In the most
successful applications pairing correlations are taken into account
in the framework of relativistic Hartree-Bogoliubov (RHB) theory
with Gogny type interactions in the pp-channel \cite{RHB}. For a
recent review see Ref.~\cite{VALR.05}.

The effective interaction entering the static part of the nucleon self-energy
is in this case given by meson exchange potentials. Without the non-linear
couplings these interactions are of Yukawa type with a range determined by the
mass of the corresponding mesons and with a relativistic structure determined
by the quantum numbers of the mesons. In the simplest case of the
$\sigma\omega$-model we have%
\begin{equation}
\tilde{V}(1,2)=-g_{\sigma}^{2}D_{\sigma}(\bm{r}_{1},\bm{r}_{2})\beta
^{(1)}\beta^{(2)}+g_{\omega}^{2}D_{\omega}(\bm{r}_{1},\bm{r}_{2}%
)(1^{(1)}1^{(2)}-{\mbox{\boldmath $\alpha$}}^{(1)}{\mbox{\boldmath $\alpha$}}%
^{(2)}) \label{OBE1}%
\end{equation}
with the Dirac matrices $\beta$ and $\bm\alpha$ and with the propagator in
$r$-space%
\begin{equation}
D_{m}(\bm{r}_{1},\bm{r}_{2})=\frac{1}{4\pi}\frac{e^{-m_{m}|\bm{r}_{1}%
-\bm{r}_{2}|}}{|\bm{r}_{1}-\bm{r}_{2}|},\text{ \ \ \ \ \ \ \ \ for
\ \ }m=\sigma,\omega. \label{OBE}%
\end{equation}
More realistic applications contain in addition a $\rho$-meson and the photon.
These potentials are obtained from the Klein-Gordon equations by elimination of
the meson degrees of freedom neglecting retardation.

In the non-linear case the static Klein-Gordon equation for the $\sigma$-meson
has the form
\begin{equation}
-\Delta\sigma+U^{\prime}(\sigma)=-g_{\sigma}\left\langle \bar{\psi}%
\psi\right\rangle , \label{KG0}%
\end{equation}
with%
\begin{equation}
U^{\prime}(\sigma)=m_{\sigma}^{2}\sigma+g_{2}\sigma^{2}+g_{3}\sigma^{3}.
\end{equation}
This is a nonlinear equation, which does not allow for an analytic
solution. The numerical solution of Eq.~(\ref{KG0}) $\sigma(\bm{r})$
provides us with
the mean field of scalar type%
\begin{equation}
\tilde{\Sigma}_{s}(\bm{r})=g_{\sigma}\sigma(\bm{r)}.
\end{equation}

Relativistic RPA is obtained as the small amplitude limit of time-dependent
Hartree theory \cite{RMG.01}. In this case the relativistic single-particle
density matrix $\rho_{kk^{\prime}}$ is expanded around the static solution
$\rho_{0}$. The linearization of the Klein-Gordon equation (\ref{KG0}) leads
to%
\begin{equation}
(-\Delta+M(\bm{r}))\delta\sigma=-g_{\sigma}{\delta\rho}_{s},
\end{equation}
where ${\delta\rho}_{s}$ is the scalar density, with $M(\bm{r})$ depending on
the coordinate $\bm{r}$
\begin{equation}
M(\bm{r})=U^{\prime\prime}(\sigma_{0}(\bm{r}))=m_{\sigma}^{2}+2g_{2}\sigma
_{0}(\bm{r})+3g_{3}\sigma_{0}^{2}(\bm{r}).
\end{equation}
where $\sigma_{0}(\bm{r})$ is the $\sigma$-field of the static solutions. In
momentum space this leads to a non-local propagator, which is the solution of
the integral equation
\begin{equation}
\bm{q}^{2}D_{\sigma}(\bm{q},\bm{q}^{\prime})+%
%TCIMACRO{\dint }%
%BeginExpansion
{\displaystyle\int}
%EndExpansion
\frac{d^{3}q^{\prime\prime}}{(2\pi)^3}~%
M(\bm{q-q}^{\prime\prime})
D_{\sigma}(\bm{q}^{\prime\prime},\bm{q}^{\prime})=
(2\pi)^3\delta(\bm{q}-\bm{q}^{\prime}),
\end{equation}
where $M(\bm{q})$ is the Fourier transform of $M(\bm{r})$:%
\begin{equation}
M(\bm{q})=%
%TCIMACRO{\dint }%
%BeginExpansion
{\displaystyle\int}
%EndExpansion
d^{3}re^{-i\bm{qr}}M(\bm{r)}.
\end{equation}
The $\sigma$ part of the effective ph-interaction used in the RPA
calculations
has therefore the form%
\begin{equation}
\tilde{V}_{\sigma}^{ph}(1,2)=-g_{\sigma}^{2}\beta^{(1)}\beta^{(2)}%
%TCIMACRO{\dint }%
%BeginExpansion
{\displaystyle\int}
%EndExpansion
\frac{d^{3}qd^{3}q^{\prime}}{(2\pi)^{6}}\ e^{i(\bm{qr}_{1}\bm{-q}^{\prime
}\bm{r}_{2})}D_{\sigma}(\bm{q},\bm{q}^{\prime}) \label{Dqqp}%
\end{equation}
and similar terms for the other mesons and the photon. We use this
interaction in the solution of the relativistic RPA equation for the
calculation of the collective phonons, as well as in
Eq.~(\ref{phonon}) for the determination of the corresponding
particle-phonon vertices $\gamma_{kl}^{\mu}$ \ and in
Eq.~(\ref{respre}) for the response function $R(14,23)$.
\newpage

\section{Density dependent meson vertices}

In recent years it has been recognized that the relativistic models are by no
means fully fledged quantum field theories. Instead of that they are effective
field theories, which provide the basis of covariant density functional
theory. Renormalizability is not important. In the spirit of density
functional theory it is reasonable to abandon the non-linear couplings and to
introduce instead of that coupling parameters $g_{\sigma}(\rho(\bm{r}))$,
$g_{\omega}(\rho(\bm{r)})$ and $g_{\rho}(\rho(\bm{r)})$, which depend on the
baryon density \cite{BT.92,FL.95,FLW.95,TW.99,DD-ME1,DD-ME2}.

In this case the static self-energy contains rearrangement terms, i.e. terms
depending on the derivative of the coupling constant with respect to the
density. We find
\begin{align}
\tilde{\Sigma}_{s}(\bm{r})  &  =g_{\sigma}\sigma(\bm{r)}\\
\tilde{\Sigma}_{0}(\bm{r})  &  =g_{\omega}\omega_{0}(\bm{r)+}g_{\rho}\tau
_{3}\rho_{0}(\bm{r)+}V_{C}(\bm{r)+}\text{ }\tilde{\Sigma}^{R}(\bm{r})
\end{align}
with the rearrangement term%
\begin{equation}
\text{ }\tilde{\Sigma}^{R}(\bm{r})=g_{\sigma}^{\prime}(\rho(\bm{r))}\rho
_{s}(\bm{r)}\sigma(\bm{r)}+g_{\omega}^{\prime}(\rho(\bm{r))}\rho
(\bm{r)}\omega_{0}(\bm{r)}+g_{\rho}^{\prime}(\rho(\bm{r))}\rho_{T}%
(\bm{r)}\rho_{0}(\bm{r)},
\end{equation}
where $\rho=\rho_{p}+\rho_{n}$ and $\rho_{T}=\rho_{p}-\rho_{n}$ are the
isoscalar and isovector baryon densities and where $\rho_{0}$ is the time-like
component of the $\rho$-field. Here we have used the fact, that we have
time-reversal invariance in the ground state and that the currents vanish in
this case.

The effective ph-interaction is obtained as the derivative of the self-energy
with respect to the density operator. We find for the $\sigma$-exchange%
\begin{align}
V_{\sigma}^{ph}(1,2)  &  =-\beta^{(1)}\beta^{(2)}g_{\sigma}(1)g_{\sigma
}(2)D_{\sigma}(\bm{r}_{1},\bm{r}_{2})-I_{\sigma}(\bm{r}_{1})\delta
(\bm{r}_{1}-\bm{r}_{2})\nonumber\\
&  -\{\beta^{(1)}1^{(2)}g_{\sigma}^{{}}(1)\tilde{g}_{\sigma}(2)+1^{(1)}%
\beta^{(2)}\tilde{g}_{\sigma}(1)g_{\sigma}^{{}}(2) + 1^{(1)}1^{(2)}\tilde
{g}_{\sigma}(1)\tilde{g}_{\sigma}(2)\}D_{\sigma}(\bm{r}_{1},\bm{r}_{2})
\end{align}
with
\begin{align}
I_{\sigma}(\bm{r}_{1})=\left\{  (\beta^{(1)}1^{(2)}+1^{(1)}\beta
^{(2)})g_{\sigma}^{\prime}(1) + 1^{(1)}1^{(2)}g_{\sigma}^{\prime\prime}%
(1)\rho_{s}({\bm r}_{1})\right\}  \times\nonumber\\
\times\int d^{3}r\ g_{\sigma}(\rho(\bm{r}))D_{\sigma}(\bm{r}_{1}%
,\bm{r})\rho_{s}(\bm{r})
\end{align}
and similar terms for the $\omega$- and $\rho$-exchange. Here we have used the
abbreviations:%
\begin{align}
g_{\sigma}(1)  &  :=g_{\sigma}(\rho(\bm{r}_{1}))\\
\tilde{g}_{\sigma}(1)  &  :=g_{\sigma}^{\prime}(\rho(\bm{r}_{1}))\rho
_{s}(\bm{r}_{1}).
\end{align}

\section{Solution in momentum space}

\label{choice}First we concentrate on the static interaction
$\tilde{V}$ of Eq.~(\ref{OBE1}) neglecting for the moment the
non-linear meson coupling. We can transform the one-boson exchange
(OBE) interaction (\ref{OBE1}) to momentum
space and neglect retardation:%
\begin{equation}
\tilde{V}(1,2)=\pm%
%TCIMACRO{\dsum \limits_{m}}%
%BeginExpansion
{\displaystyle\sum\limits_{m}}
%EndExpansion
g_{m}^{2}\int\frac{d^{3}q}{(2\pi)^{3}}e^{i\mathbf{q(r}_{1}-\mathbf{r}_{2}%
)}\Gamma_{\mu}^{m}(1)D_{m}^{\mu\nu}(q)\Gamma_{\nu}^{m}(2)
\end{equation}
where the $+$sign holds for the vector mesons and the $-$sign for the scalar
mesons. The index $m$ denotes the set of mesons $m=\{\sigma,\omega,\rho,e\}$
-- the isoscalar-scalar $\sigma$-meson, the isoscalar-vector $\omega$-meson,
the isovector-vector $\rho$-meson and the photon, which carry the respective
components of the interaction. Summation is implied also over the repeated
Greek indices. $D_{m}^{\mu\nu}(q)$ is a boson propagator which is usually
taken in the RMF calculations in the simplified Yukawa form neglecting
formfactors and retardation effects:
\begin{align}
D_{m}^{{}}(q)  &  =-\frac{1}{\mathbf{q}^{2}+m_{m}^{2}},\ \ \ \ \ m=\sigma
,\label{sigprop}\\
D_{m}^{\mu\nu}(q)  &  =\frac{g^{\mu\nu}}{\mathbf{q}^{2}+m_{m}^{2}%
},\ \ \ \ \ m=\omega,\rho,e.
\end{align}
Introducing the channel index $c=(\mathbf{q},m)$, which combines the meson
index $m$ with the momentum transfer $\mathbf{q}$, we can express $\tilde{V}$
as a sum of separable terms%
\begin{equation}
\tilde{V}(1,2)=\sum\limits_{c}d_{c}Q^{(c)}(1)Q^{(c)\dagger}(2)
\end{equation}
with%
\begin{equation}
Q^{(c)}=Q(\mathbf{q},m)=\Gamma_{\mu}^{m}e^{i\mathbf{qr}}%
\end{equation}
and the propagator
\begin{equation}
d_{c}=d(\mathbf{q},m)=\pm g_{m}^{2}D_{m}^{\mu\nu}(q).
\end{equation}
The matrix elements of $\tilde{V}$ in Dirac space have the form%
\begin{align}
\tilde{V}_{k_{1}k_{4},k_{2}k_{3}}  &  =\langle k_{1}k_{4}|\tilde{V}|k_{2}%
k_{3}\rangle=\sum\limits_{c}d_{c}Q_{k_{1}k_{2}}^{(c)}Q_{k_{3}k_{4}}^{(c)\ast
}\nonumber\\
%%%\tilde{V}_{k_{1}k_{4},k_{2}k_{3}}  &  =\langle k_{1}k_{4}|\tilde{V}|k_{2}%
%%%k_{3}\rangle=\sum\limits_{c}d_{c}Q_{k_{2}k_{1}}^{(c)\ast}Q_{k_{4}k_{3}}^{(c)}
%%%\nonumber\\
&  =\pm\sum\limits_{m}g_{m}^{2}\int\frac{d^{3}q}{(2\pi)^{3}}\langle
k_{1}|\Gamma_{\mu}^{m}e^{i\mathbf{q}\mathbf{r}}|k_{2}\rangle D_{m}^{\mu\nu
}(q)\langle k_{4}|\Gamma_{\nu}^{m}e^{-i\mathbf{q}\mathbf{r}}|k_{3}\rangle.
\label{C7}%
\end{align}
For non-linear meson-couplings this expression has to be somewhat modified,
because the propagator in momentum space (\ref{Dqqp}) is non-diagonal in this
case. We have $D_{m}^{\mu\nu}(\mathbf{q},\mathbf{q}^{\prime})$ and the values
$d_{c}$ have to be replaced by matrices $d_{cc^{\prime}}$ in this case.

\section{Response formalism in spherical nuclei}

In the spherical case angular momentum coupling reduces the numerical effort
considerably. The response equation for $\delta\rho$ reads in this case:
\begin{align}
\delta\rho_{(k_{1}k_{2})}^{J}(\omega)  &  ={\tilde{\delta\rho}}_{(k_{1}k_{2}%
)}^{J}(\omega)+\nonumber\\
&  +{\tilde{R}}_{(k_{1}k_{2})}(\omega)\sum\limits_{(k_{3}k_{4})}%
\Bigl[{\tilde{V}}_{(k_{1}k_{4},k_{2}k_{3})}^{J}+\delta\Phi_{(k_{1}k_{4}%
,k_{2}k_{3})}^{J}(\omega)\Bigr]\delta\rho_{(k_{3}k_{4})}^{J}(\omega).
\label{drho}%
\end{align}
It contains the matrix elements of the static effective interaction $\tilde
{V}^{J}$ $\ $and $\delta\Phi^{J}(\omega)$ is the change of the effective
interaction due to the energy dependence of the particle-vibration coupling.

For the meson exchange potentials we obtain after coupling to good angular
momentum $J$:%

\begin{align}
\tilde{V}_{(k_{1}k_{4},k_{2}k_{3})}^{J}  &  =\pm\frac{(4\pi)^{2}}{2J+1}%
\sum\limits_{LSm}(-1)^{S}\,\int\limits_{0}^{\infty}\frac{q^{2}dq}{(2\pi)^{3}%
}\langle k_{1}\Vert j_{L}(qr)[\Gamma_{S}^{m}Y_{L}]^{J}\Vert k_{2}\rangle
\times\nonumber\\
&  \quad\quad\quad\quad\quad\quad\quad\quad\ \ \ \ \ \ \ \ \ \ \quad\quad
\quad\quad\times\frac{g_{m}^{2}}{q^{2}+m_{m}^{2}}\langle k_{3}\Vert
j_{L}(qr)[\Gamma_{S}^{m}Y_{L}]^{J}\Vert k_{4}\rangle, \label{separable}%
\end{align}
where index $S=(0,1)$ denotes the spherical component of the Pauli matrix
entering the vertices (\ref{gammas}). The factor $(-)^{S}$ indicates that the
space-like parts of the interaction (current-current interactions) have the
opposite sign as the time-like part. The matrix elements (\ref{separable}) are
a sum of separable terms. The dimension of the matrices to be inverted is
given by the number of separable terms.

The interaction $\tilde{V}$ is treated in the response equation
Eq.~(\ref{drho4}). After a transformation of this equation to
momentum space we find
\begin{align}
\delta\rho_{LS}^{Jm}(q,\omega)  &  ={{\delta\rho}}_{LS}^{(e)Jm}(q;\omega
)\pm\,\nonumber\\
&  \pm\frac{(4\pi)^{2}}{2J+1}\sum\limits_{L^{\prime}S^{\prime}m^{\prime}%
}(-1)^{S^{\prime}}\int\limits_{0}^{\infty}\frac{q^{\prime\ 2}dq^{\prime}%
}{(2\pi)^{3}}R_{LS,L^{\prime}S^{\prime}}^{(e)Jmm^{\prime}}(q,q^{\prime}%
;\omega)\frac{g_{m^{\prime}}^{2}}{q^{\prime\ 2}+m_{m^{\prime}}^{2}}\delta
\rho_{L^{\prime}S^{\prime}}^{Jm^{\prime}}(q^{\prime};\omega), \label{drhoq}%
\end{align}
where
\begin{equation}
R_{LS,L^{\prime}S^{\prime}}^{(e)Jmm^{\prime}}(q,q^{\prime};\omega
)=\sum\limits_{(k_{1}k_{2}k_{3}k_{4})}Q_{LS(k_{1}k_{2})}^{Jm}(q)R_{(k_{1}%
k_{4},k_{2}k_{3})}^{(e)J}(\omega)Q_{L^{\prime}S^{\prime}(k_{3}k_{4}%
)}^{Jm^{\prime}}(q^{\prime})
\end{equation}
with
\begin{equation}
Q_{LS(k_{1}k_{2})}^{Jm}(q)=\langle k_{1}\Vert j_{L}(qr)[\Gamma_{S}^{m}%
Y_{L}]^{J}\Vert k_{2}\rangle
\end{equation}
is a Fourier transform of the correlated propagator $R^{(e)J}$ which
is determined in the Dirac basis by the angular momentum coupled
version of Eq.~(\ref{RE1}) :
\begin{equation}
R_{(k_{1}k_{4},k_{2}k_{3})}^{(e)J}(\omega)={\tilde{R}}_{(k_{1}k_{2})}%
(\omega)\delta_{(k_{1}k_{3})}\delta_{(k_{2}k_{4})}\,+{\tilde{R}}_{(k_{1}%
k_{2})}(\omega)\sum\limits_{(k_{5}k_{6})}\delta\Phi_{(k_{1}k_{6},k_{2}k_{5}%
)}^{J}(\omega)R_{(k_{5}k_{4},k_{6}k_{3})}^{(e)J}(\omega).
\label{correlated-propagator}%
\end{equation}
The free term on the right side of the Eq.~(\ref{drhoq}) has the
following form:
\begin{equation}
{{\delta\rho}}_{LS}^{(e)Jm}(q;\omega)=\sum\limits_{(k_{1}k_{2}k_{3}k_{4}%
)}Q_{LS(k_{1}k_{2})}^{Jm}(q)R_{(k_{1}k_{4},k_{2}k_{3})}^{(e)}(\omega
)P_{(k_{3}k_{4})}^{J},
\end{equation}
where
\begin{equation}
P_{(k_{1}k_{2})}^{J}=\langle k_{1}\Vert P^J\Vert k_{2}\rangle .
\end{equation}
The spectrum of nuclear excitations in the external field $P$ with the
multipolarity $J$ is therefore determined by the strength function:
\begin{equation}
S^{J}(E)=-\frac{1}{\pi}\lim\limits_{\Delta\rightarrow+0}Im\ \Pi^{J}%
(E+i\Delta), \label{strfq}%
\end{equation}
expressed through the following polarizability $\Pi^{J}(\omega)$:
\begin{equation}
\Pi^{J}(\omega)=\Pi^{(e)J}(\omega)\pm\frac{(4\pi)^{2}}{2J+1}\sum
\limits_{LSm}(-1)^{S}\int\limits_{0}^{\infty}\frac{dqq^{2}}{(2\pi)^{3}}%
{\delta\rho}_{LS}^{(e)Jm}(q;\omega)\frac{g_{m}^{2}}{q^{2}+m_{m}^{2}}\delta
\rho_{LS}^{Jm}(q;\omega),
\end{equation}
where
\begin{equation}
\Pi^{(e)J}(\omega)=\sum\limits_{(k_{1}k_{2}k_{3}k_{4})}P_{(k_{1}k_{2})}%
^{J}R_{(k_{1}k_{4},k_{2}k_{3})}^{(e)J}(\omega)P_{(k_{3}k_{4})}^{J} .
\end{equation}

Thus, one can see that it is convenient to solve our problem as well
as the RRPA problem in the momentum space since the rank of vectors
and matrices entering the Eq.~(\ref{drhoq}) is determined by the
number of mesh points in $q$-space and the number of meson channels.
In particular it does not depend on the mass number of the nucleus.
In realistic calculations the rank of the Eq.~(\ref{drhoq}) in the
momentum space is around 500 and, obviously, the numerical effort
does not depend considerably on the total dimension of $ph$ and
$\alpha h$ subspaces. However, if one stays in the Dirac basis the
latter dimension is exactly the rank of arrays in the
Eq.~(\ref{drho}). One should keep in mind that in practice both
subspaces are truncated at some energy differences
$\varepsilon_{ph}$ and $\varepsilon_{\alpha h}$ which are large
enough so that a further increase of these values does not influence
the results. In light nuclei with relatively small level density the
total dimension of $ph$ and $\alpha h$ subspaces is comparable and
could be even smaller than the rank of the Eq.~(\ref{drhoq}) and,
therefore, working in the Dirac basis is more preferable. For heavy
nuclei the Dirac basis becomes huge and, therefore, using of the
momentum space is more justified.

However, even if one considers the Eq.~(\ref{drhoq}) in the momentum
representation, Eq.~(\ref{correlated-propagator}) for the correlated
propagator, nevertheless, has to be solved in the Dirac basis. This
equation contains the particle-phonon coupling amplitude $\Phi^{J}$
with the following matrix elements coupled to angular momentum $J$:
\begin{align}
\Phi_{(ph^{\prime},hp^{\prime})}^{J}(\omega)  &  =\sum\limits_{\mu
}\Bigl[\delta_{(pp^{\prime})}\frac{\delta_{{\varkappa}_{h^{\prime}}{\varkappa
}_{h^{{}}}}}{2j_{h}+1}\sum\limits_{h^{\prime\prime}}\frac{\gamma
_{(h^{\prime\prime}h)}^{\mu}\gamma_{(h^{\prime\prime}h^{\prime})}^{\mu\ast}%
}{\omega-\varepsilon_{p}+\varepsilon_{h^{\prime\prime}}-\Omega^{\mu}%
}\nonumber\\
&  +\delta_{(hh^{\prime})}\frac{\delta_{{\varkappa}_{p^{\prime}}{\varkappa
}_{p^{{}}}}}{2j_{p}+1}\Bigl(\sum\limits_{p^{\prime\prime}}\frac{\gamma
_{(pp^{\prime\prime})}^{\mu}\gamma_{(p^{\prime}p^{\prime\prime})}^{\mu\ast}%
}{\omega-\varepsilon_{p^{\prime\prime}}+\varepsilon_{h}-\Omega^{\mu}}%
+\sum\limits_{{\alpha}^{\prime\prime}}\frac{\gamma_{(p{\alpha}^{\prime\prime
})}^{\mu}\gamma_{(p^{\prime}{\alpha}^{\prime\prime})}^{\mu\ast}}%
{\omega-\varepsilon_{{\alpha}^{\prime\prime}}+\varepsilon_{h}-\Omega^{\mu}%
}\Bigr)\nonumber\\
&  +(-1)^{J+J_{\mu}}\left\{
\begin{array}
[c]{ccc}%
j_{p} & j_{h} & J\\
j_{h^{\prime}} & j_{p^{\prime}} & J^{\mu}%
\end{array}
\right\}  \Bigl(\frac{(-1)^{j_{p^{\prime}}-j_{h}}\gamma_{(pp^{\prime})}^{\mu
}\gamma_{(hh^{\prime})}^{\mu\ast}}{\omega-\varepsilon_{p^{\prime}}%
+\varepsilon_{h}-\Omega^{\mu}}+\frac{(-1)^{j_{p}-j_{h^{\prime}}}%
\gamma_{(p^{\prime}p)}^{\mu\ast}\gamma_{(h^{\prime}h)}^{\mu}}{\omega
-\varepsilon_{p}+\varepsilon_{h^{\prime}}-\Omega^{\mu}}\Bigr)\Bigr],
\label{phiphc}%
\end{align}
where $\varkappa_{k}$ denotes the relativistic quantum number set:
$\varkappa_{k}=(2j_{k}+1)(l_{k}-j_{k})$, and $\gamma_{(k_{1}k_{2})}^{\mu
}=\langle k_{1}\Vert\gamma^{\mu}\Vert k_{2}\rangle$ denotes the reduced matrix
element of the particle-phonon coupling amplitude. The backwards going
components are found through the symmetry relations:
\begin{equation}
\Phi_{(hp^{\prime},ph^{\prime})}^{J}%
(\omega)=(-1)^{j_{h}+j_{p}+j_{h^{\prime}}+j_{p^{\prime}}}%
\Phi_{(p^{\prime}h,h^{\prime}p)}^{J}(-\omega).
\end{equation}
But this problem is too expensive numerically to be solved in the
full Dirac basis. Due to the pole structure of the $\Phi$ amplitude
it is naturally to suggest that particle-phonon coupling effect is
not important quantitatively far from the Fermi surface. So, in the
numerical calculations an energy window $\varepsilon_{win}$ was
implied around the Fermi surface with respect to $ph$ energy
differences $\varepsilon_{ph}$ so that the summation in the
Eq.~(\ref{correlated-propagator}) is performed only among the $ph$
pairs with $\varepsilon_{ph}\leq\varepsilon_{win}$. Consequently,
the correlated propagator differs from the mean field propagator
only within this window. This approximation has been investigated
numerically by direct calculations with different
%%%window values as
values of $\varepsilon_{win}$ as
it is discussed in Sec. \ref{results}. It is important to emphasize
that many $ph$ and $\alpha h$ configurations outside of the window
are taken into account on the RRPA level that is necessary in order
to obtain the reasonable centroid positions of giant resonances as
well as to find the dipole spurious state close to zero energy.

\bigskip

\bigskip
%==========================================================================
\leftline{\bf REFERENCES}
%\bibliographystyle{unsrt}
%\bibliographystyle{c:/a00/prsty}
%\bibliography{c:/a00/refring}
%==========================================================================

\end{document}